\documentclass{CUP-JNL-DTM}%

\usepackage{graphicx}
\usepackage{subcaption}
\usepackage{caption}
\usepackage{multicol,multirow}
\usepackage{rotating}
\usepackage{array}
\usepackage{geometry}

\usepackage{tabularx} %

\usepackage{appendix}
\usepackage[sort&compress,numbers]{natbib}
\usepackage{ifpdf}
\usepackage[T1]{fontenc}
\usepackage[utf8]{inputenc}
\usepackage{textcomp}
\usepackage{xcolor}
\definecolor{latent-purple}{RGB}{89,85,215}
\usepackage{nicematrix}
\usepackage{seqsplit}

\usepackage{amsthm} %
\usepackage{amsmath,amssymb,amsfonts}
\usepackage{newtxtext}
\usepackage{newtxmath}
\usepackage{mathrsfs}

\theoremstyle{definition}

\numberwithin{equation}{section}
\usepackage{diagbox}
\usepackage{float}
\usepackage{rotating}

\usepackage{tocloft}
\usepackage[colorlinks,allcolors=latent-purple]{hyperref}
\usepackage{cleveref}
\crefname{appendix}{App.}{Apps.}
\crefname{section}{Sec.}{Secs.}
\crefname{subsection}{Sec.}{Secs.}
\crefname{subsubsection}{Sec.}{Secs.}
\crefname{figure}{Fig.}{Figs.}
\crefname{table}{Tab.}{Tabs.}
\crefname{equation}{Eq.}{Eqs.}
\crefname{algorithm}{Alg.}{Algs.}
\crefname{example}{Ex.}{Exs.}
\crefname{page}{p.}{pp.}
\crefname{line}{l.}{ll.}

\usepackage{siunitx}
\DeclareSIUnit\angstrom{\text{Å}}

\usepackage[acronym]{glossaries}
\makeglossaries
\glsdisablehyper
\newacronym{pdb}{PDB}{Protein Data Bank}
\newacronym{afdb}{AFDB}{AlphaFold Protein Structure Database}
\newacronym{cdr}{CDR}{complementarity determining region}
\newacronym{mdm2}{MDM2}{mouse double minute 2 homolog}
\newacronym{mcl1}{MCL-1}{myeloid cell \mbox{leukemia-1}}
\newacronym{pdl1}{PD-L1}{programmed death-ligand 1}
\newacronym{il7ra}{IL-7R$\alpha$}{interleukin-7 \mbox{receptor-$\alpha$}}
\newacronym{trka}{TrkA}{tropomyosin receptor kinase A}
\newacronym{sc2rbd}{SC2RBD}{SARS-CoV-2 spike protein receptor-binding domain}
\newacronym{mDisplay}{mDisplay}{mammalian display}
\newacronym{SPR}{SPR}{surface plasmon resonance}
\newacronym{BLI}{BLI}{bio-layer interferometry}
\newacronym{MFI}{MFI}{mean fluorescence intensity}
\newacronym{HT-BLI}{HT-BLI}{high-throughput bio-layer interferometry}
\newacronym{rmsd}{RMSD}{root mean square deviation}
\newacronym{complexrmsd}{\texttt{complex\_rmsd}}{root mean squared deviation between the designed and predicted complex structures}
\newacronym{minipae}{\texttt{min\_ipae}}{minimum value across all interchain terms in the predicted alignment error matrix}
\newacronym{ipae}{\texttt{ipae}}{average across all interchain terms in the predicted alignment error matrix}
\newacronym{iptm}{\texttt{iptm}}{chain pair interface-predicted TM-score}
\newacronym{ptm}{\texttt{ptm}}{predicted TM-score}
\newacronym{ptmbinder}{\texttt{ptm\_binder}}{predicted TM-score for the binder chain}

\usepackage{fancyhdr}
\fancyheadoffset{0pt}
\newcommand{\papertitle}{\themodel: An Atom-level Frontier Model for De~Novo Protein Binder Design}
\usepackage[version=4]{mhchem}
\usepackage{siunitx}

\newcommand{\themodel}{{Latent-X}}
\newcommand{\ap}{AlphaProteo}
\newcommand{\rfdiff}{RFdiffusion}
\newcommand{\rfpep}{RFpeptides}
\newcommand{\af}[1]{AlphaFold~#1}
\newcommand{\chai}{Chai-1}
\newcommand{\boltz}{Boltz-2}

\newcommand{\denovo}{\textit{de novo}}
\newcommand{\insilico}{\textit{in silico}}

\newcommand{\ca}{\ce{C_{\alpha}}}
\newcommand{\kd}{\ensuremath{\mathrm{K}_{\mathrm{D}}}}
\newcommand{\published}[1]{\textcolor{gray}{#1}}
\newcommand{\spr}{\textsuperscript{\scalebox{1.5}{·}}}
\newcommand{\bli}{\phantom{\spr}}
\newcommand{\nobinder}{$\times$}
\newcommand{\notest}{--}

\newcommand{\subfigref}[2]{Fig.~\hyperref[#1]{\ref*{#1}#2}}

\newcommand{\customcaption}[1]{%
  \refstepcounter{figure}%
  \caption*{\textbf{Fig.~\thefigure{}~|}~#1}%
}

\newcommand{\customtablecaption}[1]{%
  \refstepcounter{table}%
  \caption*{\textbf{Tab.~\thetable{}~|}~#1}%
}

\newcommand{\ArtType}{}
\makeatletter
\newcommand{\@opjournalheader}{}
\makeatother
\newcommand{\headeright}{}

\begin{document}
\setcounter{figure}{0}
\setcounter{table}{0}

\begin{Frontmatter}
\title[Article Title]{%
    \papertitle
}

\author[]{Latent Labs Team}
\address[]{\orgaddress{\city{London, UK \& San Francisco, USA}\\
           \orgaddress{22 July 2025}} \\ 
\begin{center}
     \makebox[\textwidth][c]{\includegraphics[width=1.05\textwidth]{figures/0_summary_titlepage.pdf}}
\end{center}
\vspace{-1cm}
}

\abstract{
Traditional drug discovery relies on rounds of screening millions of candidate molecules with low success rates, making drug discovery time and resource intensive. To overcome this screening bottleneck, we introduce \themodel, an all-atom protein design model that enables a new paradigm of precision AI design. Given a target protein epitope, \themodel{} jointly generates the all atom structure and sequence of the protein binder and target, directly modelling the non-covalent interactions essential for specific binding. We demonstrate its efficacy across two therapeutically relevant modalities through extensive wet lab experiments, testing as few as 30-100 designs per target. For macrocyclic peptides, \themodel{} achieves experimental hit rates exceeding 90\% on all evaluated benchmark targets. For mini-binders, it consistently produces potent candidates against all evaluated benchmark targets, with binding affinities reaching the low nanomolar and picomolar range --- comparable to those of approved therapeutics --- whilst also being highly specific in mammalian display. In direct comparisons with the state-of-the-art models \ap, \rfdiff{} and \rfpep{} under identical conditions demonstrates, \themodel{} generates binders with higher hit rates and better binding affinities, and uniquely creates structurally diverse binders, including complex beta-sheet folds. Its end-to-end process is an order of magnitude faster than existing multi-step computational pipelines. By drastically improving the efficiency and success rate of \denovo{} design, \themodel{} represents a significant advance towards push-button biologics discovery and a valuable tool for protein engineers. \themodel{} is available at \href{https://platform.latentlabs.com}{\texttt{\color{latent-purple}https://platform.latentlabs.com}}, enabling users to reliably generate \denovo{} binders without AI infrastructure or coding.
}
\end{Frontmatter}

\section[Introduction]{Introduction}

Drug discovery faces the challenge of finding molecules that bind specifically to therapeutic targets. Traditional approaches screen millions of random compounds with success rates below 1\%, making development time-consuming and expensive. Experimental methods like phage-display and animal immunisation are resource-intensive with limitations for difficult targets and no control over binding location. Computational methods offer a compelling alternative: faster, more economical generation of precise binders without specialized domain knowledge, while unlocking new target biology and therapeutic modalities.

Protein binding enables central functions such as immune system reactions and intra- and extracellular signalling. 
As such, biological pathways can be studied and treated for therapeutic benefit by enhancing or blocking the involved interactions with precision protein binders. For this reason, the targeted creation of novel protein binders has received significant attention in both fundamental research and the development of protein therapeutics \cite{dimitrov2012therapeutic, lu2020recent, ebrahimi2023engineering}.

Depending on the application, different binder types are considered in practice. We focus on two therapeutically relevant binder modalities: macrocycles and mini-binders. Macrocycles are small cyclic peptides whose cyclization enhances degradation resistance while maintaining specificity and offering bioavailability and tissue permeability \cite{driggers2008exploration, vinogradov2019macrocyclic}. Mini-binders are short proteins of arbitrary fold that provide high binding affinity and specificity in a flexible format, representing a new therapeutic class for targeted delivery, diagnostics, and therapeutic inhibitors \cite{chevalier2017massively, cao2020novo}.

Early protein binder generation used physics-inspired energy functions \cite{basanta2020enumerative, cao2022design}. \af{2}'s success in structure prediction \cite{jumper2021highly} inspired generative models \cite{trippe2022diffusion, anand2022protein, wang2022scaffolding, ingraham2023illuminating, campbell2024generative, hayes2025simulating, chu2024all, qu2024p, lu2025all} and hallucination-based techniques \cite{anishchenko2021novo, pacesa2024bindcraft, cho2025boltzdesign1} for structure-based \denovo{} protein design. Recently, several generative models for protein binding have been published \cite{watson2023novo}, though not all are currently accessible \cite{zambaldi2024novo, chai2025chai}.

We introduce \themodel{}, a frontier all-atom model for precision protein design that jointly generates protein sequence and structure. When prompted with a target structure and hotspot residues identifying the desired epitope, \themodel{} generates \denovo{} proteins forming specific, non-covalent interactions for high-affinity binding.

We benchmark \themodel{} in wet lab experiments on mini-binders and macrocycles against the best literature-reported \ap{} \cite{zambaldi2024novo}, \rfdiff{} \cite{watson2023novo} and \rfpep{} \cite{rettie2025accurate} binders. \rfdiff{} is the most commonly used binder design method, and \rfpep{} is its macrocycle-generating variant. Both require backbone resequencing with ProteinMPNN \cite{dauparas2022robust}.

Without application-specific training, \themodel{} successfully generates both modalities, demonstrating coverage of structurally diverse binders across various lengths. The designs are structurally diverse, including beta-sheet folds that contrast with the highly helical designs of other models.

\textbf{Our main contributions are}:
\begin{enumerate}
    \item A state-of-the-art AI model for joint sequence and structure generation of diverse therapeutic binders with zero-shot \denovo{} design capability.
    \item Exceptional lab performance: 91--100\% hit rates for macrocycles and 10--64\% for mini-binders, with picomolar binding affinities and high target specificity.
    \item Universal target success: all evaluated targets bound using 30--100 tested designs per target.
    \item Superior performance over existing methods in both \insilico{} metrics and experimental validation, with 10$\times$ faster generation.
    \item End-to-end modeling of non-covalent interactions enabling epitope specificity and structurally diverse binders including complex secondary structures.
\end{enumerate}

Below we introduce the workflow and provide detailed wet lab and \insilico{} benchmarking using in-house binding assays and external affinity measurements. Access \themodel{} on \href{https://platform.latentlabs.com}{\texttt{\color{latent-purple}https://platform.latentlabs.com}}. 

\begin{figure}[H]
    \centering
    \includegraphics[width=0.99\textwidth]{figures/1_main_results.pdf}
    \customcaption{\textbf{\themodel{} generates all-atom binders with leading experimental hit rates and binding affinities.}\\ \textbf{a)} Selected lab-validated \themodel{} binders. Prompted with different targets and epitopes, \themodel{} generates diverse all-atom binder structures forming hydrogen bonds and other non-covalent interactions with the target epitope, shown in close-up structural details. Generated bonds are visualized by pink dashed lines. \textbf{b)} Experimental hit rates and binding affinities of \themodel{} binders across a range of targets. Results for macrocycles are shown in orange and for mini-binders in purple. Comparisons to prior methods are presented in \cref{tab:binding_affinities} and \cref{tab:hit_rates}.}
    \label{fig:overview}
\end{figure}

\section[Methods]{Methods}

\subsection[The \themodel{} model]{The \themodel{} model}
\begin{figure}[t]
    \centering
    \includegraphics[width=\textwidth]{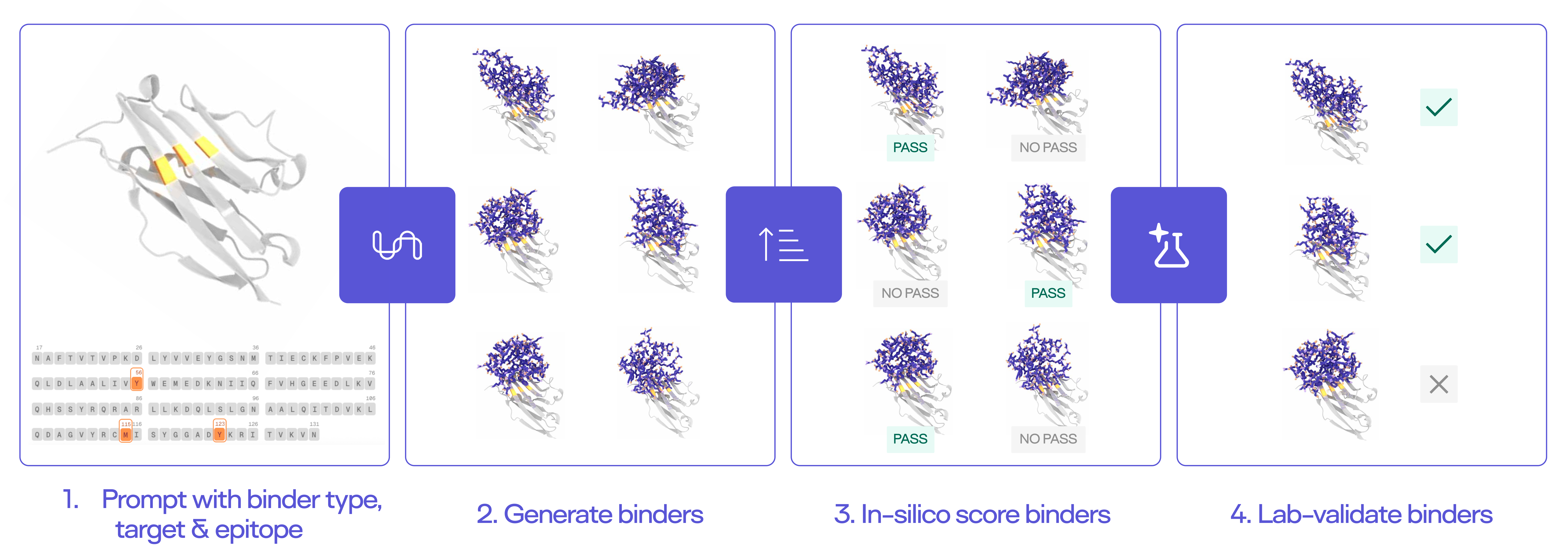}
    \customcaption{\textbf{The end-to-end binder design workflow.} Designing binders involves prompting with user inputs, generating binders with \themodel{} and scoring them with \insilico{} filters. The best designs can be selected for lab validation.}
    \label{fig:workflow}
\end{figure}

\themodel{} represents a novel approach to computational protein design that directly generates all-atom structures and sequences of binder and target proteins. The complete workflow begins with protein binder generation from \themodel{}, followed by \insilico{} filtering that allows to automatically select designs for laboratory validation, see \cref{fig:workflow}. The model can be prompted to generate requested protein binder modalities targeting specific proteins and epitopes, by providing the target protein sequence and structure, along with hotspot residue locations, see \cref{app:model_inputs}.

The model's co-generation approach for the all-atom structure of binder and target enables the formation of side chain interactions that underlie the biochemistry of binding. \themodel{} generates highly specific interactions between binder and target, including extensive hydrogen bonding networks in the binding interface, see \cref{fig:overview}. The presence and precision of these interactions represents a significant advance in protein binder design methodology that has been highly sought after in past efforts \cite{cao2022design}. By co-designing the target, the model can accommodate flexibility in the side-chain conformations and flexible loops of the target backbone where needed, whilst demonstrating an overall high fidelity to the unbound target backbone structure, see \cref{fig:target_sidechains}.

Without requiring retraining or fine-tuning, the model generalizes to binder modalities of varied topologies by producing both macrocycles and mini-binders: Mini-binders feature the open termini predominant in nature, while macrocycles have fused termini resulting in a cyclic topology. Beyond this topological diversity, \themodel{} generates functional binders with high structural diversity including complex beta sheet folds among other structural motifs, as shown in \cref{fig:dist_plot_structural_diversity}. Structural diversity is particularly important for targeting a wide range of protein surfaces and achieving specific binding interactions, but in the past, successful mini-binders have predominantly been alpha-helical bundles \cite{cao2020novo, cao2022design, bennett2023improving}.

To improve the likelihood of experimental success, the generated binders undergo automated \insilico{} filtering using established computational metrics based on structure prediction models \cite{bennett2023improving, watson2023novo, zambaldi2024novo}. The filtering process involves predicting the structure of the generated binder-target complex to determine structural self-consistency between generation and prediction. Confidence metrics predicted by structure prediction models are additionally used in the \insilico{} filters. Details are provided in \cref{app:insilico_filter}.

\themodel{} demonstrates superior computational efficiency compared to existing approaches: \themodel{} has a higher \insilico{} hit rate on unseen target structures and allows inference times that are an order of magnitude faster. This significant improvement in hit rate and inference speed enables the fast generation of experimentally relevant numbers of designs, making the approach practical for both computational experimentation within seconds and large-scale protein design applications, as detailed in \cref{app:inference_speed}.

The model's training encompasses both experimental and predicted protein structures, including single-chain and multi-chain configurations, drawing from established databases and prediction methods \cite{berman2000protein, jumper2021highly, varadi2024alphafold}. The model operates within a context window of 512 residues, for both the target and binder proteins combined. In practice this context window is suitable even when targeting proteins that exceed the context in length, since the model can generate binders to structurally cropped targets. Model details relevant to users are provided in \cref{app:model_inputs}.

Owing to continued research progress, we used two variants of the same foundational model architecture. Note that none of the variants were specifically trained or tuned for macrocycle or mini-binder design.
\begin{itemize}
\item \textbf{\themodel{} v1:} used to generate macrocycles and mini-binders for experimental validation.
\item \textbf{\themodel{} v1.1:} a slightly improved version used in \insilico{} studies of expected \insilico{} hit rates, see \cref{sec:expected_comp_hit_rate}. This is the model we initially serve on \href{https://platform.latentlabs.com}{\texttt{\color{latent-purple}https://platform.latentlabs.com}}.
\end{itemize}

\subsection[Methods for design and validation]{Methods for design and validation}

To lab-validate the performance of \themodel{}, we experimentally tested both macrocycle and mini-binder designs across seven benchmark targets that have been used to experimentally test previous state-of-the-art generative design methods \rfdiff{} \cite{watson2023novo}, \rfpep{} \cite{rettie2025accurate} and \ap{} \cite{zambaldi2024novo}. We replicated their literature-reported best-performing binders and evaluated them alongside our own designs using experimental binding measurements as detailed in \cref{app:benchmark}. This enables direct head-to-head comparison under identical lab conditions. Below, we first introduce the target proteins used and then give details on the computational and experimental workflows for each binder modality.

Our experimental evaluation focused on three key measures:
\begin{enumerate}
    \item 
\textbf{Experimental hit rate:} the proportion of designs that show measurable binding signal to their target.
\item \textbf{Binding affinity:} the interaction strength between binder and target, quantified by dissociation constant (\kd).
\item \textbf{Binding specificity:} the selectivity of binders for their intended target, assessed via cross-reactivity screening.
\end{enumerate}

All three measures have high practical and pharmacological relevance in the creation of therapeutics. High experimental hit rates reduce the number of designs that need to be lab validated, thereby reducing timelines, labour and costs. High binding affinities can directly relate to the therapeutic efficacy of a drug, for example by outcompeting other interactions, and high specificity is critical to avoid off-target effects and ensure selective function \cite{huggins2012rational}.

\subsubsection[Target proteins]{Target proteins}
\begin{figure}[h]
    \centering
    \includegraphics[width=0.9\linewidth]{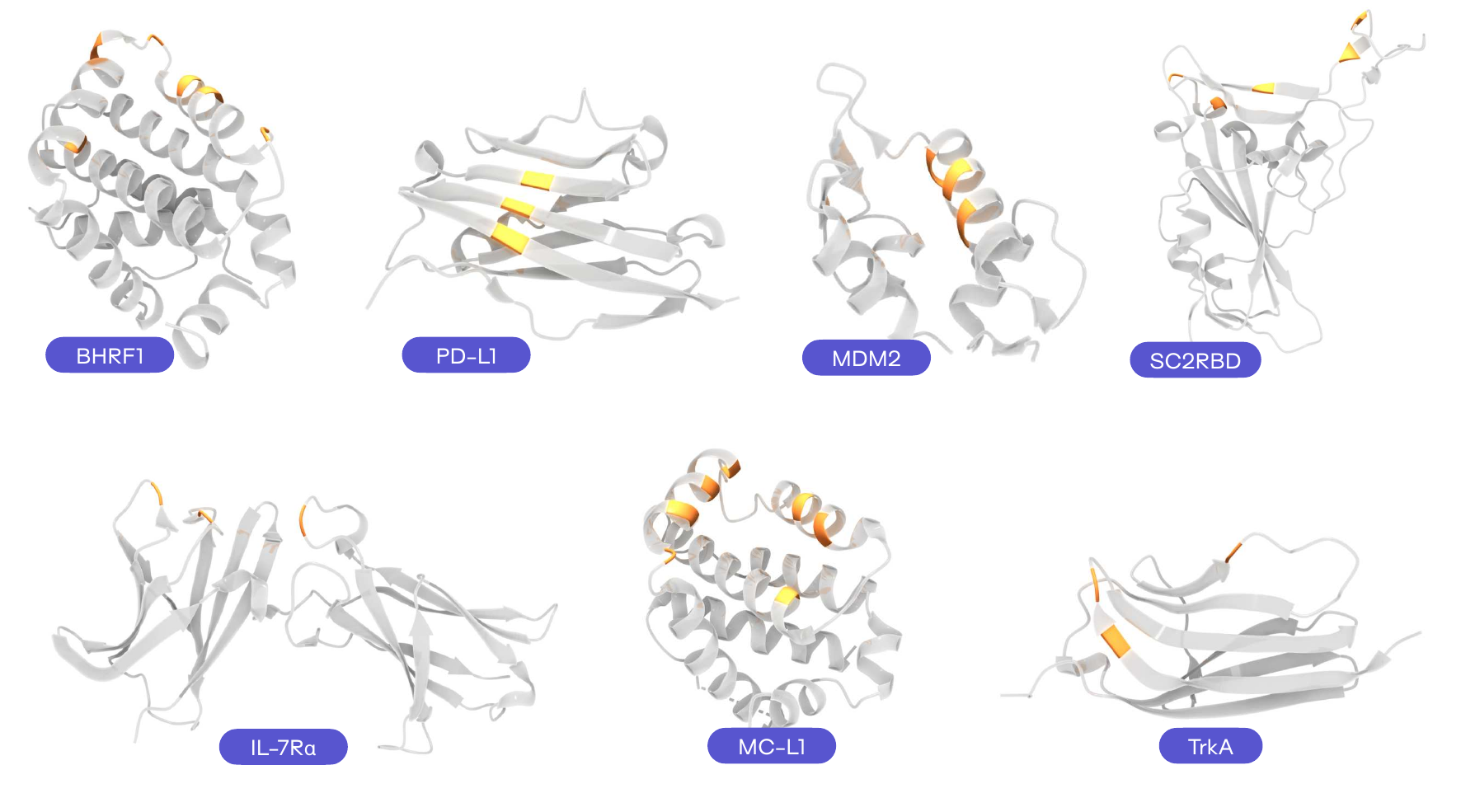}
    \customcaption{\textbf{Target proteins and hotspots used in experimental validation.} Crystal structures (grey) with hotspot residues (orange) for the seven targets used in the macrocycle and mini-binder design experiments. Further details for the targets and hotspots are provided in \cref{tab:target_table}.}
    \label{fig:Targets_w_hotspots}
\end{figure}

The benchmark targets selected for experimental validation span a range of binding difficulties and therapeutic areas, including viral entry, immune regulation, cancer, and neuropathology. We give an overview of the targets and their therapeutic relevance below, and \cref{fig:Targets_w_hotspots} visualizes the targets' crystal structures and hotspot positions.

\paragraph*{Viral infection}
\textbf{BHRF1} is an anti-apoptotic protein from Epstein–Barr virus that mimics cellular Bcl-2 proteins to suppress host cell death, facilitating viral persistence \cite{procko2014computationally}. Binders targetting an epitope on a hydrophobic groove of BHRF1 have been successfully designed before \cite{zambaldi2024novo, procko2014computationally}.

\textbf{\gls{sc2rbd}} is the receptor-binding domain of the SARS-CoV-2 spike protein and mediates viral entry via ACE2. Disrupting this interaction blocks infection, and computationally designed binders have shown promise as neutralizing agents \cite{cao2022design}.

\paragraph*{Oncology and immune regulation}
\glsreset{mdm2}
\glsreset{mcl1}
\glsreset{pdl1}
\glsreset{il7ra}
\glsreset{trka}

\textbf{\Gls{mdm2}} is a negative regulator of the tumor suppressor p53 and is frequently overexpressed in cancer \cite{shangary2008targeting}. Inhibiting the p53–MDM2 interaction offers a strategy for restoring tumor suppression.

\textbf{\Gls{mcl1}} is a anti-apoptotic Bcl-2 family member that promotes cancer cell survival and chemoresistance \cite{tantawy2023targeting}. Its helical binding groove has been previously explored in other design efforts \cite{rettie2025accurate}.

\textbf{\Gls{pdl1}} is a cell-surface immune checkpoint protein that inhibits T-cell activation and enables tumor immune evasion. Its polar and relatively flat interface makes it difficult to target with conventional binders such as antibodies \cite{gainza2023novo}.

\textbf{\Gls{il7ra}} is a cytokine receptor critical for lymphocyte development and a target in autoimmune diseases and leukemia. Its moderately hydrophobic binding interface is a tractable site for binder design, and previous work has reported variable success across different modalities \cite{berger2024preclinical}.

\paragraph*{Neuropathy}
\textbf{\Gls{trka}} is a neurotrophin receptor involved in nerve growth, pain, and inflammation \cite{mantyh2011antagonism}. Its large extracellular domain and shallow hydrophobic binding pocket pose substantial challenges for both antibody and de novo protein design \cite{cao2022design}.

\subsubsection[Macrocycle design and validation]{Macrocycle design and validation}
\label{sec:comp_design_pipeline_macro}

\paragraph{Computational design}
We \denovo{} designed macrocycles against three diverse and therapeutically relevant targets: \gls{mdm2}, \gls{mcl1} and \gls{pdl1}. Previous work using \rfpep, a variant of \rfdiff{}, successfully generated macrocycle binders against \gls{mdm2} and \gls{mcl1} \cite{rettie2025cyclic}. \gls{pdl1} has been targeted in prior rational macrocycle engineering efforts \cite{miao2021rational, fetse2022discovery}.

Hotspot residues for \gls{mdm2} and  \gls{mcl1} were determined by examining the residues engaged with the natural binding partner, as to our knowledge no prior work reports defined hotspot residues for macrocycles. For \gls{pdl1}, we use hotspot definitions previously used for mini-binder design \cite{watson2023novo}. Our hotspot definitions can be found in \cref{tab:target_table}.

For each target, we sampled 100 designs per length, for all lengths in the range 12--18, resulting in 700 designs per target. This is over an order of magnitude fewer than the numbers sampled by \rfpep{} on the same targets and epitope regions \cite{rettie2025cyclic}. Designs were scored using our macrocycle \insilico{} filter, which we describe in detail in \cref{app:insilico_filter_macrocycles}. 
The \insilico{} success rates were 67\% for \gls{mcl1}, 59\% for \gls{mdm2}, and 56\% for \gls{pdl1}. These values are also displayed in \subfigref{fig:chai_dist}{a}. 

Macrocycles can be challenging to synthesize, and previous works report high failure rate at the synthesis stage \cite{rettie2025accurate, rettie2025cyclic}. To counter this, we applied additional filtering criteria aimed at improving synthesis success. At shorter lengths, we observed recurring sequences against the same epitope, and we removed approximately 25\% of designs due to duplication. \themodel{} can generate macrocycles containing disulphide bonds, which can increase stability, as visualized in \subfigref{fig:disulphide_macrocycle}. However, due to synthesis concerns, we filter these out during sequence filtering, as described in \cref{app:computational_design_macrocycles}. To ensure our macrocycles were not a result of model memorisation, we constructed a macrocycle novelty filter using MMseqs2 \cite{steinegger2017mmseqs2}. We give further details on the synthesis and novelty pipeline in \cref{app:computational_design_macrocycles}. After applying the macrocycle structure prediction based \insilico{} filter, and filtering for novelty and synthesis feasibility, we ranked the remaining candidates by computing the \gls{ipae} and selecting the top 30 designs out of 700 for each target for experimental validation.

\paragraph{Experimental methods}
For our experiments, macrocycle designs were synthesized and cyclized using head-to-tail lactam chemistry to form the peptide bond between the termini, see \cref{app:cyclization} for more details. Since macrocycles are small proteins, they make comparatively small interfaces with their target, which can make it more challenging to achieve high binding affinities and target specificities relative to larger protein binders. Anticipating this, we used the more sensitive \gls{SPR} to assess target binding instead of \gls{BLI}.

5-point \gls{SPR} was used to perform all-against-all cross-reactivity assays to assess specificity for the top two macrocycle designs against all three macrocycle targets. For binding measurements where high precision \kd{} values were needed, 8-point \gls{SPR} was performed instead.

\subsubsection[Mini-binder design and validation]{Mini-binder design and validation}
\label{sec:comp_design_pipeline_mini}

\paragraph{Computational design}
We design novel mini-binder proteins against five diverse target proteins.  The five targets are \gls{il7ra}, \gls{pdl1}, \gls{trka}, \gls{sc2rbd} and BHRF1. \rfdiff{} was used to generate and experimentally validate mini-binders against \gls{il7ra}, \gls{pdl1}, \gls{trka} \cite{watson2023novo}, and all five proteins were used as benchmark proteins in more recent works \cite{zambaldi2024novo, chai2025chai}. We use the same \gls{pdb} entries, hotspots, and target chain and residue numbers as described in these recent works \cite{watson2023novo, zambaldi2024novo}. Further details can be found in \cref{tab:target_table}. 

For each target, we generated a generous \num{20000} designs in the length range 80--120. This provided us with sufficient samples to arrive at the intended minimum of 100 final designs per target, whilst simultaneously allowing us to test a wide variety of novel and structurally diverse binders. As an initial screen, we filtered the designs using the \insilico{} filter. The \insilico{} success rates were 50.6\% for BHRF1, 7.2\% for \gls{trka}, 11\% for \gls{pdl1}, 9.2\% for \gls{il7ra}, and 10\% for \gls{sc2rbd}. These values are also displayed in \subfigref{fig:chai_dist}{b}. 

To demonstrate \themodel{}’s ability to generate novel proteins, we discarded proteins that do not pass our novelty threshold. Furthermore, to maximize structural diversity, we clustered our designs using Foldseek \cite{van2024fast} and selected a limited number of designs belonging to each clusters. Details of the novelty threshold and structural clustering can be found in \cref{app:computational_design_minibinders}. Similar to previous works, we omitted designs containing cysteines due to complications that may come in expression and purification and the potential for intermolecular disulphide bonds forming \cite{watson2023novo, verkuil2022language}.

\paragraph{Experimental methods}
The 100 mini-binder designs for each of the five targets were evaluated using a two-tiered screening strategy. Tier 1 served as a fast pre-screen to select binders for Tier 2, wherein binding affinity was determined with high resolution. For Tier 1 testing, we used \gls{HT-BLI} to detect binding. This allowed us to calculate hit rates, defined as the fraction of designs exceeding our pre-defined binding response threshold, see \cref{app:BLI}. Tier 1 hits that showed sufficient binding signal were advanced to Tier 2, where binding affinities were quantified using 5-point \gls{BLI}, see \cref{app:BLI}. From these measurements, we computed \kd{} values, where lower values indicate stronger binding interactions.

To complement HT-BLI and support downstream therapeutic applications, we also validated our mini-binder designs using our in-house \gls{mDisplay} assay. \gls{mDisplay} is conceptually similar to yeast surface display but operates in a mammalian expression system using HEK293T cells \cite{ho2009mammalian}, with the output being binding signal derived from mean fluorescence intensity (MFI) representing amount of interaction between binder and target proteins (see \cref{fig:Definition of mDisplay} and \cref{app:mdisplay} for more details). This offers advantages when screening therapeutic candidates that benefit from native folding, disulphide bond formation, or post-translational modifications \cite{dyson2020beyond}. This dual-screening approach allowed us to cross-validate hit identification across orthogonal binding platforms. Positive correlation was observed between approaches, as shown in  \cref{fig:mammalian HT-BLI correlation}. Finally, \gls{mDisplay} was used to perform all-against-all binding assays to assess cross-target specificity among the top four binders for three of the considered mini-binder targets.

\section[Laboratory results]{Laboratory results}

We tested \themodel's ability to generate successful binders for two binder modalities of therapeutic relevance: macrocycles and mini-binders. We found that \themodel{} is able to generate \denovo{} binders for both, exceeding the binding affinities of prior methods under identical experimental conditions, hitting every target that was tested with as few as 30--100 designs per target, and displaying high target specificities.

A detailed presentation of experimental results for each binder modality is provided in the following. We show the all-atom structures of all successfully lab-validated binders generated by \themodel{} in complex with their targets in \cref{fig:macrocycle_wall_of_best_hits} and \cref{fig:minibinder_wall_of_best_hits}. For each target, we provide a selection of lab-validated \themodel{} generated mini-binder and macrocycle sequences in \cref{tab:Binder_sequences} to allow for external reproduction.

\begin{table}[h!]
\centering
\begin{tabular}{lrrrrrrrr}
    & \multicolumn{3}{c}{Macrocycles} & \multicolumn{5}{c}{Mini-binders} \\
    \cmidrule(lr){2-4} \cmidrule(lr){5-9}
    \diagbox[linewidth=0.2pt,font=\small]{Model}{Target}
     & \acrshort{mdm2} & \acrshort{mcl1} & \acrshort{pdl1} & BHRF1 & \acrshort{trka} & \acrshort{pdl1} & \acrshort{il7ra} & \acrshort{sc2rbd} \\
    \midrule
    \multicolumn{1}{r}{} & \multicolumn{3}{c}{\kd{} (\textmu{}M) $\scriptstyle\downarrow$} & \multicolumn{5}{c}{\kd{} (nM) $\scriptstyle\downarrow$} \\
    \midrule
    
    Latent-X & \textbf{5.35}\spr & 18.4\spr & \textbf{71.7}\spr & \textbf{22.5}\bli & \textbf{0.04}\bli & 0.27\bli & \textbf{\textless0.01}\bli & \textbf{\textless0.01}\bli \\
    \addlinespace[0.5ex]
    
    AlphaProteo & & & & & & & & \\
    \quad replicated & \notest\bli & \notest\bli & \notest\bli & 458.0\bli & \nobinder\bli & \textbf{\textless0.01}\bli & \textbf{\textless0.01}\bli & 0.17\bli \\
    \quad \published{published} & \published{\notest}\bli & \published{\notest}\bli & \published{\notest}\bli & \published{8.5}\bli & \published{0.96}\bli & \published{0.18}\bli & \published{0.08}\bli & \published{26}\bli \\
    \addlinespace[0.5ex]
    
    \rfdiff{}/\rfpep{} & & & & & & & & \\
    \quad replicated & 8.38\spr & \textbf{10.0}\spr & \notest\bli & \notest\bli & \nobinder\bli & 699\spr & 31.1\bli & \notest\bli \\
    \quad \published{published} & \published{1.90}\bli & \published{2.0}\bli & \published{\notest}\bli & \published{\notest}\bli & \published{328}\bli & \published{1400}\bli & \published{30.0}\bli & \published{\notest}\bli \\
    
\end{tabular}
\vspace{0.3cm}
\customtablecaption{\textbf{Binding affinities of the best binders from \themodel{}, AlphaProteo, \rfdiff{} and \rfpep.} Summary of binding affinities for the best binder of \themodel{} and best binders reported by AlphaProteo, \rfdiff{} and \rfpep{} as published and replicated in identical assays for each benchmark target. Lower affinity values (\kd) indicate tighter binding ($\scriptstyle\downarrow$) and the best binder per target is highlighted in \textbf{bold}, comparing to replicas where possible. \rfdiff{} was used for mini-binder designs, and \rfpep{} was used for macrocycle designs to ensure modality-matched comparisons. Dotted values were measured using \gls{SPR}, all other measurements were performed with \gls{BLI}. A cross (\nobinder) indicates no discernable binding was observed.}
\label{tab:binding_affinities}
\end{table}

\renewcommand{\published}[1]{#1}

\begin{table}[h!]
\centering
\begin{tabular}{lrrrrrrrr}
& \multicolumn{3}{c}{Macrocycles} & \multicolumn{5}{c}{Mini-binders} \\
\cmidrule(lr){2-4} \cmidrule(lr){5-9}
\diagbox[linewidth=0.2pt,font=\small]{Model}{Target}
 & \acrshort{mdm2} & \acrshort{mcl1} & \acrshort{pdl1} & BHRF1 & \acrshort{trka} & \acrshort{pdl1} & \acrshort{il7ra} & \acrshort{sc2rbd} \\
\midrule
\multicolumn{1}{r}{} & \multicolumn{3}{c}{Hit Rate (\%) $\scriptstyle\uparrow$} & \multicolumn{5}{c}{Hit Rate (\%) $\scriptstyle\uparrow$} \\
\midrule

\themodel{} & \textbf{90.9}\bli & \textbf{100.0}\bli & \textbf{94.1}\bli & 64.0\bli & \textbf{10.0}\bli & \textbf{49.0}\bli & 26.0\bli & \textbf{52.0}\bli \\
\addlinespace[0.5ex]

AlphaProteo & & & & & & & & \\
\quad \published{published} & \published{\notest}\bli & \published{\notest}\bli & \published{\notest}\bli & \textbf{88.0}\bli & \published{9.0}\bli & \published{15.0}\bli & \published{25.0}\bli & \published{12.0}\bli \\
\addlinespace[0.5ex]

\rfdiff{}/\rfpep{} & & & & & & & & \\
\quad \published{published} & \published{37.5}\bli & \published{21.4}\bli & \published{\notest}\bli & \published{\notest}\bli & \published{6.3}\bli & \published{12.6}\bli & \textbf{33.7}\bli & \published{\notest}\bli \\

\end{tabular}
\vspace{0.3cm}
\customtablecaption{\textbf{Experimental hit rates for binders from \themodel{}, AlphaProteo, \rfdiff{} and \rfpep{}.} Summary of hit rates for each method across benchmark targets. The best hit rate per target is highlighted in \textbf{bold}. Hit rates are calculated as the number of designs with measureable binding divided by the total number of designs tested, described in  \cref{app:BLI}. For \themodel{}, 100 mini-binders were tested per target, and 11-17 macrocycles were tested per target. Hit rates for \ap{} and \rfdiff{}/\rfpep{} are taken from the original publications respectively \cite{zambaldi2024novo, watson2023novo, rettie2025cyclic}. Note that experimental methods and thresholds for measurable binding can vary across modalities and models.}
\label{tab:hit_rates}
\end{table}

\subsection[Macrocycle laboratory results]{Macrocycle laboratory results}
\label{sec:macro_results}

For each macrocycle target, we submitted 30 designs for experimental testing. Of these, 77\% for \gls{mcl1}, 57\% for \gls{mdm2}, and 87\% for \gls{pdl1} were successfully synthesized and cyclized, with purity exceeding 90\%, see \cref{app:cyclization}. Between 11 and 17 designs per target were selected for binding assessment by \gls{SPR}, chosen to provide a representative sample of the successfully produced peptides.

\themodel{} generated successful binders for all three targets, achieving a hit rate of > 90\% across targets, see \cref{fig:overview}, \cref{fig:biophysical_characterisation}, \cref{fig:raw_sensogram_macrocycle}, and \cref{tab:hit_rates}. For \gls{mdm2} and \gls{mcl1}, our hit rates were 91\% and 100\%, respectively, exceeding those reported for \rfpep{}, which achieved detectable binding in 38\% of designs for \gls{mdm2} and 21\% for \gls{mcl1} \cite{rettie2025accurate}. In addition to the targets previously addressed by \rfpep{}, \themodel{} also successfully generated macrocycles for \gls{pdl1}. Our chosen epitope regions span diverse structural features, including helical binding grooves in \gls{mcl1} and \gls{mdm2} and a flat $\beta$-sheet interface in \gls{pdl1}, yet \themodel{} consistently produced binders across nearly all tested macrocycles.

\themodel{} generated macrocycles for \gls{mdm2} and \gls{mcl1} with binding affinities comparable to those of \rfpep{} against the same epitopes, see \cref{tab:binding_affinities}. Binding affinities for macrocycles can be improved by substituting non-canonical amino acids into designed macrocycles, as shown in \cite{rettie2025cyclic}. We used literature macrocycle positive controls as listed in \cref{tab:positive-binders}.

To assess specificity, we performed all-against-all \gls{SPR} binding experiments across the different targets, see \subfigref{fig:specificity_assay}{a}. As expected for macrocycles with no non-canonical modifications, low-level off-target binding was observed even in positive control peptides. Nevertheless, the Latent-X designs consistently showed high specificity: each macrocycle bound most strongly to its intended target, with off-target affinities typically several orders of magnitude weaker. This demonstrates that despite using only canonical amino acids, \themodel{} can generate specific macrocycles.

\begin{figure}[H]
    \centering
    \includegraphics[width=1\textwidth]{figures/4_biophysical_characterisation.pdf}
    \customcaption{\textbf{Biophysical characterization of the best \themodel{} generated macrocycles and mini-binders.} Two of the top-performing \themodel{} binders for each target and modality with their designed bound structures and binding curves. Macrocycles (orange) were measured by \gls{SPR} using eight concentrations and analysed using a steady-state model to determine \kd{}. For raw sensogram data of the macrocycles, see \cref{fig:raw_sensogram_macrocycle}. Mini-binders (purple) were assessed by \gls{BLI} using five concentrations with kinetic fitting. Reported \kd{} values span from low picomolar to micromolar affinities with lower \kd{} values corresponding to stronger binding.}
    \label{fig:biophysical_characterisation}
\end{figure}

\subsection[Mini-binder laboratory results]{Mini-binder laboratory results}
\label{sec:mini_results}

For each mini-binder target, we submitted 100 \themodel{} designs for validation via \gls{HT-BLI}. A subset of 88 designs per target (to fit 96-well plates with controls) was tested in parallel using \gls{mDisplay}. 

\themodel{} achieved high hit rates across all targets, demonstrating consistent success in generating binders with measurable target engagement, see \cref{fig:overview} and \cref{tab:hit_rates}. Hit rates as determined via  \gls{HT-BLI} ranged from 64\% on BHRF1 to 10\% on \gls{trka}. Notably, we observed substantially improved performance on several challenging targets compared to AlphaProteo, including a 52\% hit rate on \gls{sc2rbd} versus 12\%, and 49\% on \gls{pdl1} versus 15\%, representing over 4-fold and 3-fold gains, respectively \cite{zambaldi2024novo}. Note that hit rates are subject to experimental sensitivity thresholds which vary by the experimental method that is chosen, allowing only approximate comparison.

From the \gls{HT-BLI} screen, we selected a subset of hits, based on high association rates ($K_a$) and response units (RU), for binding affinity characterization by 5-point \gls{BLI}. To benchmark our performance, we replicated and tested the self-reported top designs from \rfdiff{} and \ap{}, see \cref{tab:binding_affinities}. \themodel{} generated binders with strong affinities ranging from low picomolar to low nanomolar, see \cref{fig:biophysical_characterisation}. Across all five mini-binder targets, our designs consistently outperformed or matched published and replicated designs from \ap{} and \rfdiff. On \gls{sc2rbd} and BHRF1, our best mini-binders showed an approximately 20-fold improvement over AlphaProteo, comparing to replicated binding affinities. We used literature mini-binder positive controls as listed in \cref{tab:positive-binders}.

\begin{figure}
    \centering
    \includegraphics[width=1\linewidth]{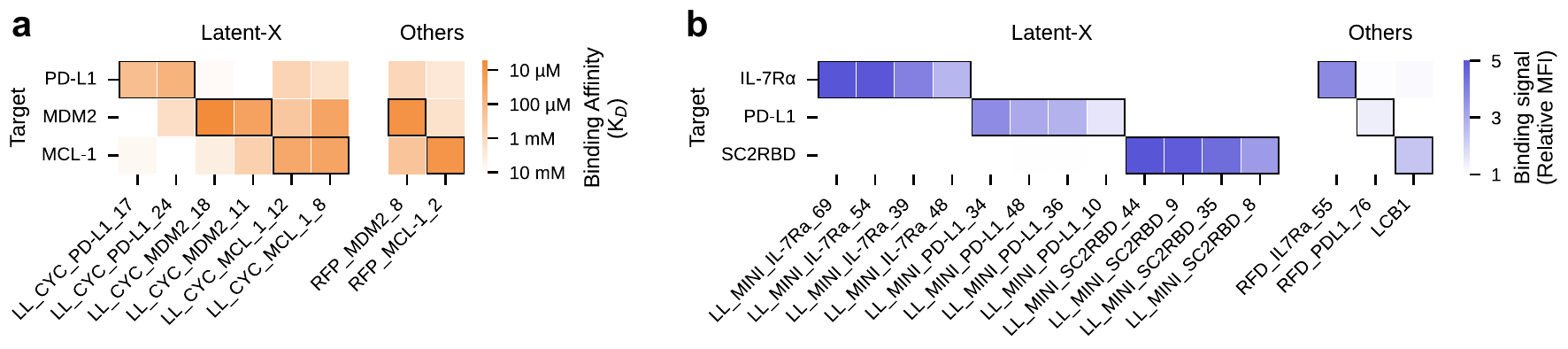}
    \customcaption{\textbf{Latent-X binders display high specificity for their intended targets.}
    Specificity was assessed with all-against-all binding experiments, with high specificity corresponding to strictly on-diagonal binding signal. The on-diagonal for each target is highlighted with a black outline. \textbf{a)} Macrocycle specificity experiments: We report the binding affinities of six \themodel{} generated macrocycles for three targets, as determined by \gls{SPR}. We additionally report the experiment's results for the two best literature-reported \rfpep{} macrocycles. \textbf{b)} Mini-binder specificity experiments: We report the binding signal of 12 \themodel{} generated mini-binders for three targets, as determined by \gls{mDisplay}. We additionally report the experiment's results for three literature-reported mini-binder designs.}
    \label{fig:specificity_assay}
\end{figure}

In \gls{mDisplay}, the dynamic range in \gls{MFI} readout was robust for three of the five tested targets (\gls{il7ra}, \gls{pdl1} and \gls{sc2rbd}), enabling quantitative discrimination of binder specificity via the \gls{mDisplay} platform. We observed good concordance between \gls{mDisplay} and HT-BLI across these targets (Pearson's \textit{r} = 0.68–0.79, as seen in \cref{fig:mammalian HT-BLI correlation}), supporting the robustness of mDisplay as a quantitative binding platform. We performed all-against-all \gls{mDisplay} binding experiments between these three targets and the top four \themodel{} binders indentified using 5-point \gls{BLI}. As shown in \subfigref{fig:specificity_assay}{b}, we found \themodel{} generated mini-binders to be highly target-specific, with no detectable off-target binding in \gls{mDisplay}.

\section{Expected \insilico{} hit rates}
\label{sec:expected_comp_hit_rate}

We assessed the fraction of binder designs that pass our \insilico{} filters across a large and diverse set of previously unseen protein targets and epitopes. This provides quantitative estimates of the \insilico{} hit rates that users can anticipate when applying the model to arbitrary protein targets and is similar to the \insilico{} benchmark conducted in \cite[][Fig. S2]{zambaldi2024novo}. This hit rate is target- and epitope-dependent, and we therefore report the distribution of \insilico{} hit rates that we found over a large held-out set of protein targets and epitopes.

\subsection[Benchmarking protocol]{Benchmarking protocol}
\label{sec:benchmarking_protocol}

We constructed a set of 200 target protein structures that were recently added to the \gls{pdb}, ensuring that the examples were not part of the training data for \themodel{}, structure prediction models used in the \insilico{} filters, or the design methods we benchmark against (i.e., \rfdiff{} and \rfpep{}).
Details of how we constructed the dataset can be found in \cref{app:expected_comp_hit_rate}. We used the same set of 200 target proteins to estimate the \insilico{} hit rate for both, macrocycles and mini-binders, and benchmarked against \rfpep{} and \rfdiff{} using the same dataset, the same hotspots and the same \insilico{} filters.

In order to simulate the targeting of different admissible epitopes, we algorithmically constructed three different sets of hotspots for each target. For the purpose of benchmarking we defined an epitope by three hotspot residues, which we automatically identify and select as detailed in \cref{app:hotspot_picker}. For all benchmarked methods, i.e. \themodel{} and \rfpep{} or \rfdiff{}, we sampled 100 binders per target and epitope for each of both the macrocycle and mini-binder task. For macrocycles we randomly sampled binder lengths in the range of 12--18 and for mini-binders in the length range 80--120 amino acids. This resulted in $200 \times{} 3 \times{} 100 = \num{60000}$ generated designs per model and binder modality. 

For each target, we define the \insilico{} hit rate as the number of binder designs that pass our \insilico{} filter for macrocycles and mini-binders respectively, averaged over all three epitopes, similar to AlphaProteo \cite{zambaldi2024novo}. The resolution of our protocol is a hit rate of 1/600. Targets for which we failed to produce binders in this setting may produce binders that pass our filter if further samples were generated. To facilitate independent verification, we used openly accessible structure prediction models. For instance, in \cref{fig:chai_dist} the \insilico{} filter used was based on \chai{}. We repeated the analysis with the \boltz{} based alternative, run on the same designs that were scored with \chai{}. The results are qualitatively similar and reported in \cref{sec:expected_comp_hit_rate_boltz}.

\subsection[Macrocyle \insilico{} hit rates]{Macrocyle \insilico{} hit rates}

\themodel{} achieved significantly higher expected \insilico{} macrocycle hit rates compared to \rfpep{}. \themodel{} showed an average hit rate of 8.26\% and hits 30\% of the targets with more than 10\% \insilico{} hit rate, compared to \rfpep{}---the \rfdiff{} macrocycle workflow---that averaged 1.72\%, hitting 1.5\% of the targets with more than 10\% success. Furthermore, \themodel{} displayed a far lower failure rate, meaning fewer targets on which \themodel{} failed to produce any binders in the sample of 600.

In \subfigref{fig:chai_dist}{a} we show the distribution of hit rates across the test set of 200 \gls{pdb} targets. The computational success rates for the designed macrocycles generated for experimental validation, see \cref{sec:comp_design_pipeline_macro}, is highlighted with grey dashed lines. For these targets we observe a success rate higher than any randomly selected target and epitope, indicating that the \insilico{} hit rate can be improved even further by a judicious choice of suitable hotspots.  

\begin{figure}[htb]
    \centering
    \includegraphics[width=0.9\linewidth]{figures/6_distplot_chai.pdf}
    \customcaption{\textbf{\textit{In silico} hit rates for \themodel{} vs \rfpep{} / \rfdiff{}.} \textbf{a)} Distribution of \insilico{} hit rates for macrocycles on 200 held-out targets. \themodel{}'s hit rates are shown in orange, and those of \rfpep{} in grey. \textbf{b)} Distribution of \insilico{} hit rates for mini-binders on 200 held-out targets. \themodel{}'s hit rates are shown in purple, and those of \rfdiff{} in grey. Distributions show the binned relative frequency against the y-axis shown on the left. Curves show the cumulative distribution functions against the y-axis on the right. Dashed vertical lines show the \insilico{} success rates achieved in design workflows for lab validation, see \cref{sec:comp_design_pipeline_macro} and \cref{sec:comp_design_pipeline_mini}. \textit{In silico} filter based on \chai{}.}
    \label{fig:chai_dist}
\end{figure}

\subsection[Mini-binder \insilico{} hit rates]{Mini-binder \insilico{} hit rates}
For mini-binders, \themodel{} compared favourably to \rfdiff{} with an average \insilico{} hit rate per target of 5.11\% compared to 3.02\%. Similarly, \themodel{} achieved \insilico{} hit rates of more than 10\% for more than 14\% of targets, compared with 5\% of targets for \rfdiff{}. For \themodel{} we were unable to find binders above our detection threshold for $5.5\%$ of all targets. \rfdiff{} on the other hand failed to generate binders for $9.5\%$ of all targets. The distribution of hit rates over all 200 targets is shown in \subfigref{fig:chai_dist}{b}. The \insilico{} hit rates for the designed mini-binders generated for experimental validation are annotated with grey dashed lines.

In \subfigref{fig:chai_dist}{b} we show the computational success rates for the designed mini-binders generated for experimental validation with grey dashed lines.

\section[Discussion]{Discussion}
\label{sec:discussion}

\subsection[Key findings and implications]{Key findings and implications}
This work presents a significant advancement in computational protein design through the development of \themodel{}, a frontier AI model that transforms the process of protein binder design from random screening to precision engineering. Our key findings demonstrate that joint generation of protein sequence and structure can achieve unprecedented success rates in creating functional protein binders.

The experimental validation reveals breakthrough results with \themodel{} achieving 91--100\% hit rates for macrocycles and 10--64\% for mini-binders, representing substantial improvements over existing methods. The binding affinities achieved---ranging from picomolar to low micromolar---demonstrate that computationally designed binders can match or exceed the performance of traditional discovery methods while requiring orders of magnitude fewer experimental tests.

The structural diversity of generated binders represents another significant finding. Unlike existing methods that predominantly produce helical designs, \themodel{} generates diverse folds including beta-sheet structures, expanding the accessible design space for protein therapeutics. This diversity is particularly important for targeting challenging protein surfaces that may not be amenable to conventional binding motifs.

With inference times an order of magnitude faster than existing models and higher \insilico{} hit rates, \themodel{} makes large-scale protein design computationally feasible. This efficiency, combined with the high experimental success rates, has profound implications for drug discovery, potentially replacing traditional approaches that screen millions of compounds with $<1\%$ success rates through targeted design approaches generating successful binders in batches of 30--100 designs.

\subsection[Comparison with existing methods]{Comparison with existing methods}
Our head-to-head experimental comparison under identical laboratory conditions provides definitive evidence of \themodel{}'s leading performance. Against \ap{}, we observed substantial improvements including 4-fold higher hit rates on \gls{sc2rbd} (52\% vs. 12\%) and 3-fold improvements on \gls{pdl1} (49\% vs. 15\%), with binding affinity improvements of approximately 20-fold on both \gls{sc2rbd} and BHRF1.

Compared to \rfdiff{} and \rfpep{}, the most widely used existing methods, \themodel{} demonstrates consistent advantages across all tested targets. For macrocycles, our hit rates of 91--100\% substantially exceed \rfpep{}'s reported 21--38\% success rates on the same targets. The computational benchmarking on 200 diverse \gls{pdb} targets further confirms these advantages, with \themodel{} achieving 8.26\% vs 1.72\% \insilico{} hit rates for macrocycles and 5.11\% vs 3.02\% for mini-binders compared to baseline methods.

A key architectural difference enabling these improvements is \themodel{}'s joint generation of sequence and structure. Existing methods typically use sequential pipelines where backbone generation and sequence design are performed independently, potentially missing optimal sequence-structure combinations. \themodel{}'s co-generation approach allows simultaneous optimization of both components, combined with the ability to directly generate non-covalent bonds across the protein-protein interface. Rather than relying on post-hoc optimization, \themodel{} learns the biochemistry required for specific binding interactions at the atomistic level, resulting in more realistic and functional designs.

\subsection[Limitations and future work]{Limitations and future work}
Despite these advances, several limitations warrant consideration for future development. The generated proteins can exhibit sequence similarity to natural proteins, which may limit novelty in certain applications. The design process currently assumes access to reasonable quality target structures, whether experimentally determined or computationally predicted. This requirement may limit applicability to targets lacking structural information, though the rapid advancement in structure prediction methods continues to expand the accessible target space.

The current model operates within a 512 amino acid context window including the target protein. While not a limiting factor in most practical binding applications, this may constrain rare applications where very large binders are desired.

The \insilico{} filters used for design selection were optimized based on lab-validated mini-binders of 45--65 amino acids with predominantly helical folds. This bias may introduce selection artifacts in the form of false positives and negatives, especially at longer lengths and for non-helical structures. Future work should expand the \insilico{} filter tuning data to include diverse binder lengths and structural motifs.

Several promising directions for future development emerge from this work. Expanding to additional binder modalities beyond macrocycles and mini-binders could demonstrate broader applicability. Developing approaches for multi-target binders or binders with designed specificity profiles could enable more sophisticated therapeutic applications. Integration with additional biochemical constraints such as stability, expression levels, and pharmacokinetic properties directly into the generation process rather than as post-hoc filters could yield more drug-like designs.

\section[Conclusion]{Conclusion}
This work represents a significant milestone in computational protein design, demonstrating that AI-driven approaches can achieve unprecedented success in generating functional protein binders. \themodel{}'s combination of joint sequence-structure generation, all-atom precision, and superior experimental performance establishes a new standard for the field.

The experimental validation across seven therapeutically relevant targets and two distinct binder modalities provides robust evidence that computational protein design has reached a level of maturity suitable for practical drug discovery applications. The ability to generate high-affinity binders with high hit rates and structural diversity addresses longstanding challenges in protein therapeutics development.

The availability of \themodel{} through the online platform democratizes advanced protein design capabilities, allowing researchers without specialized AI infrastructure to access state-of-the-art design tools. This accessibility could accelerate discoveries across the broader scientific community.

Looking forward, the continued development of AI-driven protein design methods promises to further transform therapeutic development. As these approaches mature and expand to encompass additional modalities and constraints, they may ultimately enable the routine generation of protein therapeutics tailored to specific targets and applications. The success demonstrated in this work provides a foundation for continued advancement, bringing us closer to predictable, efficient, and broadly accessible protein therapeutic development.

\section*{Contributors}
Alex Bridgland, Jonathan Crabbé, Henry Kenlay, Daniella Pretorius, Sebastian M. Schmon, Agrin Hilmkil, Rebecca Bartke-Croughan, Robin Rombach\textsuperscript{**}, Michael Flashman\textsuperscript{*}, Tomas Matteson, Simon Mathis, Alexander W. R. Nelson\textsuperscript{**}, David Yuan, Annette Obika, Simon A. A. Kohl\textsuperscript{***}

\textsuperscript{***} Corresponding author. E-mail: \href{mailto:simon@latentlabs.com}{\texttt{simon@latentlabs.com}}.\\
\phantom{*}\textsuperscript{**} Work performed as an advisor to Latent Labs.\\
\phantom{**}\textsuperscript{*} Work performed while at Latent Labs.\\

\paragraph{Author contributions}

\textbf{Conceptualization and team leadership:} S.K. conceived the research direction, led the team, and with A.B. developed the initial machine learning codebase. D.Y., D.P., S.K. contributed to experimental design. A.O., A.N., contributed to project delivery and narrative.

\textbf{Machine learning development:} S.K., J.C., A.B., S.S., H.K., A.H. developed the ML pre-training methodology and the model architecture. A.B., S.K., J.C., H.K. built the data pipeline. J.C., H.K., S.S., S.K. developed post-training methods. R.R. provided guidance. A.B., A.H., J.C., H.K. contributed to ML engineering.

\textbf{Computational design and evaluation:} H.K., D.P., S.K. led computational protein design workflows. D.P., A.B., M.F., R.B.C. performed computational evaluation of lab results. S.S., H.K. performed computational benchmarking studies.

\textbf{Experimental validation:} D.Y., R.B.C. conducted all in-house wet lab experiments. D.Y., A.O., D.P. managed external laboratory partnerships.

\textbf{Model serving via web platform:} A.H. led the technical development of the web platform and managed outsourced work, with contributions from M.F., A.B. and S.S. A.H. built the model API and inference pipeline, J.C. contributed to API monitoring. S.K. conceived of the web platform. A.O., S.K. oversaw delivery of written standards and documented commitments.

\textbf{Writing and review:} S.K. conceived of the manuscript structure, and S.K., D.P., S.S., H.K., A.B., D.Y., R.B.C., J.C. contributed to writing and figure making. S.M., T.M. provided manuscript review and feedback.

All authors contributed to the work and approved the final manuscript.

\paragraph{Competing interests}
All authors have contributed as employees of or advisors to Latent Labs Technologies Inc. or Latent Labs Limited. 

\paragraph{Acknowledgments}
We thank Mária Vlachynská for her organizational support, creative input on our visual materials, and for orchestrating the collaborative spaces and celebratory moments that brought this work to fruition. We also extend our gratitude to Krishan Bhatt for his thoughtful contributions to our documentation and communication materials. 

\bibliographystyle{unsrtnat}
\bibliography{references}

\begin{thebibliography}{55}
\providecommand{\natexlab}[1]{#1}
\providecommand{\url}[1]{\texttt{#1}}
\expandafter\ifx\csname urlstyle\endcsname\relax
  \providecommand{\doi}[1]{doi: #1}\else
  \providecommand{\doi}{doi: \begingroup \urlstyle{rm}\Url}\fi

\bibitem[Dimitrov(2012)]{dimitrov2012therapeutic}
Dimiter~S Dimitrov.
\newblock Therapeutic proteins.
\newblock \emph{Therapeutic Proteins: Methods and Protocols}, pages 1--26, 2012.

\bibitem[Lu et~al.(2020)Lu, Zhou, He, Jiang, Peng, Tong, and Shi]{lu2020recent}
Haiying Lu, Qiaodan Zhou, Jun He, Zhongliang Jiang, Cheng Peng, Rongsheng Tong, and Jianyou Shi.
\newblock Recent advances in the development of protein--protein interactions modulators: mechanisms and clinical trials.
\newblock \emph{Signal transduction and targeted therapy}, 5\penalty0 (1):\penalty0 213, 2020.

\bibitem[Ebrahimi and Samanta(2023)]{ebrahimi2023engineering}
Sasha~B Ebrahimi and Devleena Samanta.
\newblock Engineering protein-based therapeutics through structural and chemical design.
\newblock \emph{Nature communications}, 14\penalty0 (1):\penalty0 2411, 2023.

\bibitem[Driggers et~al.(2008)Driggers, Hale, Lee, and Terrett]{driggers2008exploration}
Edward~M Driggers, Stephen~P Hale, Jinbo Lee, and Nicholas~K Terrett.
\newblock The exploration of macrocycles for drug discovery—an underexploited structural class.
\newblock \emph{Nature Reviews Drug Discovery}, 7\penalty0 (7):\penalty0 608--624, 2008.

\bibitem[Vinogradov et~al.(2019)Vinogradov, Yin, and Suga]{vinogradov2019macrocyclic}
Alexander~A Vinogradov, Yizhen Yin, and Hiroaki Suga.
\newblock Macrocyclic peptides as drug candidates: {R}ecent progress and remaining challenges.
\newblock \emph{Journal of the American Chemical Society}, 141\penalty0 (10):\penalty0 4167--4181, 2019.

\bibitem[Chevalier et~al.(2017)Chevalier, Silva, Rocklin, Hicks, Vergara, Murapa, Bernard, Zhang, Lam, Yao, et~al.]{chevalier2017massively}
Aaron Chevalier, Daniel-Adriano Silva, Gabriel~J Rocklin, Derrick~R Hicks, Renan Vergara, Patience Murapa, Steffen~M Bernard, Lu~Zhang, Kwok-Ho Lam, Guorui Yao, et~al.
\newblock Massively parallel de novo protein design for targeted therapeutics.
\newblock \emph{Nature}, 550\penalty0 (7674):\penalty0 74--79, 2017.

\bibitem[Cao et~al.(2020)Cao, Goreshnik, Coventry, Case, Miller, Kozodoy, Chen, Carter, Walls, Park, et~al.]{cao2020novo}
Longxing Cao, Inna Goreshnik, Brian Coventry, James~Brett Case, Lauren Miller, Lisa Kozodoy, Rita~E Chen, Lauren Carter, Alexandra~C Walls, Young-Jun Park, et~al.
\newblock {De novo design of picomolar {SARS-C}o{V}-2 miniprotein inhibitors}.
\newblock \emph{Science}, 370\penalty0 (6515):\penalty0 426--431, 2020.

\bibitem[Basanta et~al.(2020)Basanta, Bick, Bera, Norn, Chow, Carter, Goreshnik, Dimaio, and Baker]{basanta2020enumerative}
Benjamin Basanta, Matthew~J Bick, Asim~K Bera, Christoffer Norn, Cameron~M Chow, Lauren~P Carter, Inna Goreshnik, Frank Dimaio, and David Baker.
\newblock An enumerative algorithm for de novo design of proteins with diverse pocket structures.
\newblock \emph{Proceedings of the National Academy of Sciences}, 117\penalty0 (36):\penalty0 22135--22145, 2020.

\bibitem[Cao et~al.(2022)Cao, Coventry, Goreshnik, Huang, Sheffler, Park, Jude, Markovi{\'c}, Kadam, Verschueren, et~al.]{cao2022design}
Longxing Cao, Brian Coventry, Inna Goreshnik, Buwei Huang, William Sheffler, Joon~Sung Park, Kevin~M Jude, Iva Markovi{\'c}, Rameshwar~U Kadam, Koen~HG Verschueren, et~al.
\newblock Design of protein-binding proteins from the target structure alone.
\newblock \emph{Nature}, 605\penalty0 (7910):\penalty0 551--560, 2022.

\bibitem[Jumper et~al.(2021)Jumper, Evans, Pritzel, Green, Figurnov, Ronneberger, Tunyasuvunakool, Bates, {\v{Z}}{\'\i}dek, Potapenko, et~al.]{jumper2021highly}
John Jumper, Richard Evans, Alexander Pritzel, Tim Green, Michael Figurnov, Olaf Ronneberger, Kathryn Tunyasuvunakool, Russ Bates, Augustin {\v{Z}}{\'\i}dek, Anna Potapenko, et~al.
\newblock Highly accurate protein structure prediction with {A}lpha{F}old.
\newblock \emph{nature}, 596\penalty0 (7873):\penalty0 583--589, 2021.

\bibitem[Trippe et~al.(2022)Trippe, Yim, Tischer, Baker, Broderick, Barzilay, and Jaakkola]{trippe2022diffusion}
Brian~L Trippe, Jason Yim, Doug Tischer, David Baker, Tamara Broderick, Regina Barzilay, and Tommi Jaakkola.
\newblock Diffusion probabilistic modeling of protein backbones in {3D} for the motif-scaffolding problem.
\newblock \emph{arXiv preprint arXiv:2206.04119}, 2022.

\bibitem[Anand and Achim(2022)]{anand2022protein}
Namrata Anand and Tudor Achim.
\newblock Protein structure and sequence generation with equivariant denoising diffusion probabilistic models.
\newblock \emph{arXiv preprint arXiv:2205.15019}, 2022.

\bibitem[Wang et~al.(2022)Wang, Lisanza, Juergens, Tischer, Watson, Castro, Ragotte, Saragovi, Milles, Baek, et~al.]{wang2022scaffolding}
Jue Wang, Sidney Lisanza, David Juergens, Doug Tischer, Joseph~L Watson, Karla~M Castro, Robert Ragotte, Amijai Saragovi, Lukas~F Milles, Minkyung Baek, et~al.
\newblock Scaffolding protein functional sites using deep learning.
\newblock \emph{Science}, 377\penalty0 (6604):\penalty0 387--394, 2022.

\bibitem[Ingraham et~al.(2023)Ingraham, Baranov, Costello, Barber, Wang, Ismail, Frappier, Lord, Ng-Thow-Hing, Van~Vlack, et~al.]{ingraham2023illuminating}
John~B Ingraham, Max Baranov, Zak Costello, Karl~W Barber, Wujie Wang, Ahmed Ismail, Vincent Frappier, Dana~M Lord, Christopher Ng-Thow-Hing, Erik~R Van~Vlack, et~al.
\newblock Illuminating protein space with a programmable generative model.
\newblock \emph{Nature}, 623\penalty0 (7989):\penalty0 1070--1078, 2023.

\bibitem[Campbell et~al.(2024)Campbell, Yim, Barzilay, Rainforth, and Jaakkola]{campbell2024generative}
Andrew Campbell, Jason Yim, Regina Barzilay, Tom Rainforth, and Tommi Jaakkola.
\newblock Generative flows on discrete state-spaces: {E}nabling multimodal flows with applications to protein co-design.
\newblock \emph{arXiv preprint arXiv:2402.04997}, 2024.

\bibitem[Hayes et~al.(2025)Hayes, Rao, Akin, Sofroniew, Oktay, Lin, Verkuil, Tran, Deaton, Wiggert, et~al.]{hayes2025simulating}
Thomas Hayes, Roshan Rao, Halil Akin, Nicholas~J Sofroniew, Deniz Oktay, Zeming Lin, Robert Verkuil, Vincent~Q Tran, Jonathan Deaton, Marius Wiggert, et~al.
\newblock Simulating 500 million years of evolution with a language model.
\newblock \emph{Science}, page eads0018, 2025.

\bibitem[Chu et~al.(2024)Chu, Kim, Cheng, El~Nesr, Xu, Shuai, and Huang]{chu2024all}
Alexander~E Chu, Jinho Kim, Lucy Cheng, Gina El~Nesr, Minkai Xu, Richard~W Shuai, and Po-Ssu Huang.
\newblock An all-atom protein generative model.
\newblock \emph{Proceedings of the National Academy of Sciences}, 121\penalty0 (27):\penalty0 e2311500121, 2024.

\bibitem[Qu et~al.(2024)Qu, Guan, Ma, Zhai, Wu, and Wang]{qu2024p}
Wei Qu, Jiawei Guan, Rui Ma, Ke~Zhai, Weikun Wu, and Haobo Wang.
\newblock P(all-atom) is unlocking new path for protein design.
\newblock \emph{bioRxiv}, pages 2024--08, 2024.

\bibitem[Lu et~al.(2025)Lu, Yan, Robinson, Kelow, Yang, Gligorijevic, Cho, Bonneau, Abbeel, and Frey]{lu2025all}
Amy~X Lu, Wilson Yan, Sarah~A Robinson, Simon Kelow, Kevin~K Yang, Vladimir Gligorijevic, Kyunghyun Cho, Richard Bonneau, Pieter Abbeel, and Nathan~C Frey.
\newblock All-atom protein generation with latent diffusion.
\newblock In \emph{ICLR 2025 Workshop on Generative and Experimental Perspectives for Biomolecular Design}, 2025.

\bibitem[Anishchenko et~al.(2021)Anishchenko, Pellock, Chidyausiku, Ramelot, Ovchinnikov, Hao, Bafna, Norn, Kang, Bera, et~al.]{anishchenko2021novo}
Ivan Anishchenko, Samuel~J Pellock, Tamuka~M Chidyausiku, Theresa~A Ramelot, Sergey Ovchinnikov, Jingzhou Hao, Khushboo Bafna, Christoffer Norn, Alex Kang, Asim~K Bera, et~al.
\newblock De novo protein design by deep network hallucination.
\newblock \emph{Nature}, 600\penalty0 (7889):\penalty0 547--552, 2021.

\bibitem[Pacesa et~al.(2024)Pacesa, Nickel, Schellhaas, Schmidt, Pyatova, Kissling, Barendse, Choudhury, Kapoor, Alcaraz-Serna, et~al.]{pacesa2024bindcraft}
Martin Pacesa, Lennart Nickel, Christian Schellhaas, Joseph Schmidt, Ekaterina Pyatova, Lucas Kissling, Patrick Barendse, Jagrity Choudhury, Srajan Kapoor, Ana Alcaraz-Serna, et~al.
\newblock {BindCraft}: one-shot design of functional protein binders.
\newblock \emph{bioRxiv}, pages 2024--09, 2024.

\bibitem[Cho et~al.(2025)Cho, Pacesa, Zhang, Correia, and Ovchinnikov]{cho2025boltzdesign1}
Yehlin Cho, Martin Pacesa, Zhidian Zhang, Bruno~E Correia, and Sergey Ovchinnikov.
\newblock Boltzdesign1: {I}nverting all-atom structure prediction model for generalized biomolecular binder design.
\newblock \emph{bioRxiv}, pages 2025--04, 2025.

\bibitem[Watson et~al.(2023)Watson, Juergens, Bennett, Trippe, Yim, Eisenach, Ahern, Borst, Ragotte, Milles, et~al.]{watson2023novo}
Joseph~L Watson, David Juergens, Nathaniel~R Bennett, Brian~L Trippe, Jason Yim, Helen~E Eisenach, Woody Ahern, Andrew~J Borst, Robert~J Ragotte, Lukas~F Milles, et~al.
\newblock De novo design of protein structure and function with {RFdiffusion}.
\newblock \emph{Nature}, 620\penalty0 (7976):\penalty0 1089--1100, 2023.

\bibitem[Zambaldi et~al.(2024)Zambaldi, La, Chu, Patani, Danson, Kwan, Frerix, Schneider, Saxton, Thillaisundaram, et~al.]{zambaldi2024novo}
Vinicius Zambaldi, David La, Alexander~E Chu, Harshnira Patani, Amy~E Danson, Tristan~OC Kwan, Thomas Frerix, Rosalia~G Schneider, David Saxton, Ashok Thillaisundaram, et~al.
\newblock De novo design of high-affinity protein binders with {AlphaProteo}.
\newblock \emph{arXiv preprint arXiv:2409.08022}, 2024.

\bibitem[Team et~al.(2025)Team, Boitreaud, Dent, Geisz, McPartlon, Meier, Qiao, Rogozhnikov, Rollins, Wollenhaupt, and Wu]{chai2025chai}
Chai~Discovery Team, Jacques Boitreaud, Jack Dent, Danny Geisz, Matthew McPartlon, Joshua Meier, Zhuoran Qiao, Alex Rogozhnikov, Nathan Rollins, Paul Wollenhaupt, and Kevin Wu.
\newblock Zero‑shot antibody design in a 24‑well plate.
\newblock \emph{bioRxiv}, 2025.

\bibitem[Rettie et~al.(2025{\natexlab{a}})Rettie, Juergens, Adebomi, Bueso, Zhao, Leveille, Liu, Bera, Wilms, {\"U}ffing, et~al.]{rettie2025accurate}
Stephen~A Rettie, David Juergens, Victor Adebomi, Yensi~Flores Bueso, Qinqin Zhao, Alexandria~N Leveille, Andi Liu, Asim~K Bera, Joana~A Wilms, Alina {\"U}ffing, et~al.
\newblock Accurate de novo design of high-affinity protein-binding macrocycles using deep learning.
\newblock \emph{Nature Chemical Biology}, pages 1--9, 2025{\natexlab{a}}.

\bibitem[Dauparas et~al.(2022)Dauparas, Anishchenko, Bennett, Bai, Ragotte, Milles, Wicky, Courbet, de~Haas, Bethel, et~al.]{dauparas2022robust}
Justas Dauparas, Ivan Anishchenko, Nathaniel Bennett, Hua Bai, Robert~J Ragotte, Lukas~F Milles, Basile~IM Wicky, Alexis Courbet, Rob~J de~Haas, Neville Bethel, et~al.
\newblock Robust deep learning--based protein sequence design using {ProteinMPNN}.
\newblock \emph{Science}, 378\penalty0 (6615):\penalty0 49--56, 2022.

\bibitem[Bennett et~al.(2023)Bennett, Coventry, Goreshnik, Huang, Allen, Vafeados, Peng, Dauparas, Baek, Stewart, et~al.]{bennett2023improving}
Nathaniel~R Bennett, Brian Coventry, Inna Goreshnik, Buwei Huang, Aza Allen, Dionne Vafeados, Ying~Po Peng, Justas Dauparas, Minkyung Baek, Lance Stewart, et~al.
\newblock Improving de novo protein binder design with deep learning.
\newblock \emph{Nature Communications}, 14\penalty0 (1):\penalty0 2625, 2023.

\bibitem[Berman et~al.(2000)Berman, Westbrook, Feng, Gilliland, Bhat, Weissig, Shindyalov, and Bourne]{berman2000protein}
Helen~M Berman, John Westbrook, Zukang Feng, Gary Gilliland, Talapady~N Bhat, Helge Weissig, Ilya~N Shindyalov, and Philip~E Bourne.
\newblock The protein data bank.
\newblock \emph{Nucleic acids research}, 28\penalty0 (1):\penalty0 235--242, 2000.

\bibitem[Varadi et~al.(2024)Varadi, Bertoni, Magana, Paramval, Pidruchna, Radhakrishnan, Tsenkov, Nair, Mirdita, Yeo, et~al.]{varadi2024alphafold}
Mihaly Varadi, Damian Bertoni, Paulyna Magana, Urmila Paramval, Ivanna Pidruchna, Malarvizhi Radhakrishnan, Maxim Tsenkov, Sreenath Nair, Milot Mirdita, Jingi Yeo, et~al.
\newblock {AlphaFold} protein structure database in 2024: {P}roviding structure coverage for over 214 million protein sequences.
\newblock \emph{Nucleic acids research}, 52\penalty0 (D1):\penalty0 D368--D375, 2024.

\bibitem[Huggins et~al.(2012)Huggins, Sherman, and Tidor]{huggins2012rational}
David~J Huggins, Woody Sherman, and Bruce Tidor.
\newblock Rational approaches to improving selectivity in drug design.
\newblock \emph{Journal of medicinal chemistry}, 55\penalty0 (4):\penalty0 1424--1444, 2012.

\bibitem[Procko et~al.(2014)Procko, Berguig, Shen, Song, Frayo, Convertine, Margineantu, Booth, Correia, Cheng, et~al.]{procko2014computationally}
Erik Procko, Geoffrey~Y Berguig, Betty~W Shen, Yifan Song, Shani Frayo, Anthony~J Convertine, Daciana Margineantu, Garrett Booth, Bruno~E Correia, Yuanhua Cheng, et~al.
\newblock A computationally designed inhibitor of an {Epstein-Barr viral Bcl-2} protein induces apoptosis in infected cells.
\newblock \emph{Cell}, 157\penalty0 (7):\penalty0 1644--1656, 2014.

\bibitem[Shangary and Wang(2008)]{shangary2008targeting}
Sanjeev Shangary and Shaomeng Wang.
\newblock Targeting the {MDM2-p53} interaction for cancer therapy.
\newblock \emph{Clinical Cancer Research}, 14\penalty0 (17):\penalty0 5318--5324, 2008.

\bibitem[Tantawy et~al.(2023)Tantawy, Timofeeva, Sarkar, and Gandhi]{tantawy2023targeting}
Shady~I Tantawy, Natalia Timofeeva, Aloke Sarkar, and Varsha Gandhi.
\newblock Targeting {MCL-1} protein to treat cancer: opportunities and challenges.
\newblock \emph{Frontiers in oncology}, 13:\penalty0 1226289, 2023.

\bibitem[Gainza et~al.(2023)Gainza, Wehrle, Van Hall-Beauvais, Marchand, Scheck, Harteveld, Buckley, Ni, Tan, Sverrisson, et~al.]{gainza2023novo}
Pablo Gainza, Sarah Wehrle, Alexandra Van Hall-Beauvais, Anthony Marchand, Andreas Scheck, Zander Harteveld, Stephen Buckley, Dongchun Ni, Shuguang Tan, Freyr Sverrisson, et~al.
\newblock De novo design of protein interactions with learned surface fingerprints.
\newblock \emph{Nature}, 617\penalty0 (7959):\penalty0 176--184, 2023.

\bibitem[Berger et~al.(2024)Berger, Seeger, Yu, Aydin, Yang, Rosenblum, Guenin-Mac{\'e}, Glassman, Arguinchona, Sniezek, et~al.]{berger2024preclinical}
Stephanie Berger, Franziska Seeger, Ta-Yi Yu, Merve Aydin, Huilin Yang, Daniel Rosenblum, Laure Guenin-Mac{\'e}, Caleb Glassman, Lauren Arguinchona, Catherine Sniezek, et~al.
\newblock Preclinical proof of principle for orally delivered {Th17} antagonist miniproteins.
\newblock \emph{Cell}, 187\penalty0 (16):\penalty0 4305--4317, 2024.

\bibitem[Mantyh et~al.(2011)Mantyh, Koltzenburg, Mendell, Tive, and Shelton]{mantyh2011antagonism}
Patrick~W Mantyh, Martin Koltzenburg, Lorne~M Mendell, Leslie Tive, and David~L Shelton.
\newblock Antagonism of nerve growth {factor-TrkA} signaling and the relief of pain.
\newblock \emph{Anesthesiology}, 115\penalty0 (1):\penalty0 189, 2011.

\bibitem[Rettie et~al.(2025{\natexlab{b}})Rettie, Campbell, Bera, Kang, Kozlov, Bueso, De~La~Cruz, Ahlrichs, Cheng, Gerben, et~al.]{rettie2025cyclic}
Stephen~A Rettie, Katelyn~V Campbell, Asim~K Bera, Alex Kang, Simon Kozlov, Yensi~Flores Bueso, Joshmyn De~La~Cruz, Maggie Ahlrichs, Suna Cheng, Stacey~R Gerben, et~al.
\newblock Cyclic peptide structure prediction and design using {AlphaFold2}.
\newblock \emph{Nature Communications}, 16\penalty0 (1):\penalty0 1--15, 2025{\natexlab{b}}.

\bibitem[Miao et~al.(2021)Miao, Zhang, Zhang, Li, Zhu, and Jiang]{miao2021rational}
Qi~Miao, Wanheng Zhang, Kuojun Zhang, He~Li, Jidong Zhu, and Sheng Jiang.
\newblock {Rational design of a potent macrocyclic peptide inhibitor targeting the PD-1/PD-L1 protein--protein interaction}.
\newblock \emph{RSC advances}, 11\penalty0 (38):\penalty0 23270--23279, 2021.

\bibitem[Fetse et~al.(2022)Fetse, Zhao, Liu, Mamani, Mustafa, Adhikary, Ibrahim, Liu, Patel, Nakhjiri, et~al.]{fetse2022discovery}
John Fetse, Zhen Zhao, Hao Liu, Umar-Farouk Mamani, Bahaa Mustafa, Pratik Adhikary, Mohammed Ibrahim, Yanli Liu, Pratikkumar Patel, Maryam Nakhjiri, et~al.
\newblock Discovery of cyclic peptide inhibitors targeting pd-l1 for cancer immunotherapy.
\newblock \emph{Journal of medicinal chemistry}, 65\penalty0 (18):\penalty0 12002--12013, 2022.

\bibitem[Steinegger and S{\"o}ding(2017)]{steinegger2017mmseqs2}
Martin Steinegger and Johannes S{\"o}ding.
\newblock {MMseqs2} enables sensitive protein sequence searching for the analysis of massive data sets.
\newblock \emph{Nature biotechnology}, 35\penalty0 (11):\penalty0 1026--1028, 2017.

\bibitem[Van~Kempen et~al.(2024)Van~Kempen, Kim, Tumescheit, Mirdita, Lee, Gilchrist, S{\"o}ding, and Steinegger]{van2024fast}
Michel Van~Kempen, Stephanie~S Kim, Charlotte Tumescheit, Milot Mirdita, Jeongjae Lee, Cameron~LM Gilchrist, Johannes S{\"o}ding, and Martin Steinegger.
\newblock Fast and accurate protein structure search with foldseek.
\newblock \emph{Nature biotechnology}, 42\penalty0 (2):\penalty0 243--246, 2024.

\bibitem[Verkuil et~al.(2022)Verkuil, Kabeli, Du, Wicky, Milles, Dauparas, Baker, Ovchinnikov, Sercu, and Rives]{verkuil2022language}
Robert Verkuil, Ori Kabeli, Yilun Du, Basile~IM Wicky, Lukas~F Milles, Justas Dauparas, David Baker, Sergey Ovchinnikov, Tom Sercu, and Alexander Rives.
\newblock Language models generalize beyond natural proteins.
\newblock \emph{BioRxiv}, pages 2022--12, 2022.

\bibitem[Ho and Pastan(2009)]{ho2009mammalian}
Mitchell Ho and Ira Pastan.
\newblock Mammalian cell display for antibody engineering.
\newblock \emph{Methods in Molecular Biology}, 525:\penalty0 337--352, xiv, 2009.

\bibitem[Dyson et~al.(2020)Dyson, Masters, Pazeraitis, Perera, Syrjanen, Surade, Thorsteinson, Parthiban, Jones, Sattar, et~al.]{dyson2020beyond}
Michael~R Dyson, Edward Masters, Deividas Pazeraitis, Rajika~L Perera, Johanna~L Syrjanen, Sachin Surade, Nels Thorsteinson, Kothai Parthiban, Philip~C Jones, Maheen Sattar, et~al.
\newblock Beyond affinity: {S}election of antibody variants with optimal biophysical properties and reduced immunogenicity from mammalian display libraries.
\newblock In \emph{MAbs}, volume~12, page 1829335. Taylor \& Francis, 2020.

\bibitem[Bourne et~al.(1997)Bourne, Berman, McMahon, Watenpaugh, Westbrook, and Fitzgerald]{bourne199730}
Philip~E Bourne, Helen~M Berman, Brian McMahon, Keith~D Watenpaugh, John~D Westbrook, and Paula~MD Fitzgerald.
\newblock Macromolecular crystallographic information file {(mmCIF)}.
\newblock In \emph{Methods in enzymology}, volume 277, pages 571--590. Elsevier, 1997.

\bibitem[Mariani et~al.(2013)Mariani, Biasini, Barbato, and Schwede]{mariani2013lddt}
Valerio Mariani, Marco Biasini, Alessandro Barbato, and Torsten Schwede.
\newblock {lDDT}: {a} local superposition-free score for comparing protein structures and models using distance difference tests.
\newblock \emph{Bioinformatics}, 29\penalty0 (21):\penalty0 2722--2728, 2013.

\bibitem[Abramson et~al.(2024)Abramson, Adler, Dunger, Evans, Green, Pritzel, Ronneberger, Willmore, Ballard, Bambrick, et~al.]{abramson2024accurate}
Josh Abramson, Jonas Adler, Jack Dunger, Richard Evans, Tim Green, Alexander Pritzel, Olaf Ronneberger, Lindsay Willmore, Andrew~J Ballard, Joshua Bambrick, et~al.
\newblock Accurate structure prediction of biomolecular interactions with {AlphaFold} 3.
\newblock \emph{Nature}, pages 1--3, 2024.

\bibitem[team et~al.(2024)team, Boitreaud, Dent, McPartlon, Meier, Reis, Rogozhonikov, and Wu]{chai2024chai}
Chai~Discovery team, Jacques Boitreaud, Jack Dent, Matthew McPartlon, Joshua Meier, Vinicius Reis, Alex Rogozhonikov, and Kevin Wu.
\newblock Chai-1: {D}ecoding the molecular interactions of life.
\newblock \emph{BioRxiv}, 2024.

\bibitem[Passaro et~al.(2025)Passaro, Corso, Wohlwend, Reveiz, Thaler, Ram~Somnath, Getz, Portnoi, Roy, Stark, et~al.]{passaro2025boltz}
Saro Passaro, Gabriele Corso, Jeremy Wohlwend, Mateo Reveiz, Stephan Thaler, Vignesh Ram~Somnath, Noah Getz, Tally Portnoi, Julien Roy, Hannes Stark, et~al.
\newblock Boltz-2: {T}owards accurate and efficient binding affinity prediction.
\newblock \emph{BioRxiv}, pages 2025--06, 2025.

\bibitem[Xu and Zhang(2010)]{xu2010significant}
Jinrui Xu and Yang Zhang.
\newblock How significant is a protein structure similarity with {TM-score}= 0.5?
\newblock \emph{Bioinformatics}, 26\penalty0 (7):\penalty0 889--895, 2010.

\bibitem[Suzek et~al.(2015)Suzek, Wang, Huang, McGarvey, Wu, and Consortium]{suzek2015uniref}
Baris~E Suzek, Yuqi Wang, Hongzhan Huang, Peter~B McGarvey, Cathy~H Wu, and UniProt Consortium.
\newblock Uniref clusters: {A} comprehensive and scalable alternative for improving sequence similarity searches.
\newblock \emph{Bioinformatics}, 31\penalty0 (6):\penalty0 926--932, 2015.

\bibitem[Cock et~al.(2009)Cock, Antao, Chang, Chapman, Cox, Dalke, Friedberg, Hamelryck, Kauff, Wilczynski, et~al.]{cock2009biopython}
Peter~JA Cock, Tiago Antao, Jeffrey~T Chang, Brad~A Chapman, Cymon~J Cox, Andrew Dalke, Iddo Friedberg, Thomas Hamelryck, Frank Kauff, Bartek Wilczynski, et~al.
\newblock Biopython: {f}reely available python tools for computational molecular biology and bioinformatics.
\newblock \emph{Bioinformatics}, 25\penalty0 (11):\penalty0 1422--1423, 2009.

\bibitem[Kabsch and Sander(1983)]{kabsch1983dictionary}
Wolfgang Kabsch and Christian Sander.
\newblock Dictionary of protein secondary structure: pattern recognition of hydrogen-bonded and geometrical features.
\newblock \emph{Biopolymers: Original Research on Biomolecules}, 22\penalty0 (12):\penalty0 2577--2637, 1983.

\bibitem[{Twist Bioscience}(2025)]{twist_codon_tool}
{Twist Bioscience}.
\newblock Codon optimization tool.
\newblock \url{https://www.twistbioscience.com/resources/digital-tools/codon-optimization-tool}, 2025.
\newblock Accessed: 2025-07-16.

\end{thebibliography}

\clearpage
\appendix
\thispagestyle{empty} %
\section*{\Titlefont Supplementary information}

\renewcommand{\thefigure}{S\arabic{figure}}
\renewcommand{\theHfigure}{S\arabic{figure}}
\setcounter{figure}{0}
\renewcommand{\thetable}{S\arabic{table}}
\setcounter{table}{0}

\section[Model details for users]{Model details for users}
\themodel{} is available at \href{https://platform.latentlabs.com}{\texttt{\color{latent-purple}https://platform.latentlabs.com}}. In the following we provide relevant details on model inputs, training, and model outputs that allow users to reason about model usage.

\subsection[Model training and evaluation]{Model training and evaluation}

We trained \themodel{} on data from a mixture of the \gls{pdb}~\cite{berman2000protein} and version 4 of the \gls{afdb}~\cite{varadi2024alphafold} with a context window of 512 residues.

For the \gls{pdb} dataset, we imposed a date cut-off that excludes any entry added after the 23 November 2023. We also excluded any entry resolved with resolution over \SI{5}{\angstrom} or determined with nuclear magnetic resonance (NMR) spectroscopy. When processing \gls{pdb} structures, we dropped any residue that is structurally unresolved. If several biological assemblies are provided in an mmCIF file \cite{bourne199730}, we generated the one appearing first in the file. 

\subsection[Model inputs]{Model inputs}
\label{app:model_inputs}

\themodel{} requires the following user provided inputs:
\begin{enumerate}
    \item \textbf{Target structure:} The mmCIF file that describes the structure and sequence of the target protein(s) for which binders are designed. The target can consist of multiple protein chains or crops of protein chains.
    \item \textbf{Hotspot residues:} The sequence location of the subset of target residues that constitute the target's binding hotspots. At least one hotspot needs to be provided. In practice a small number of surface accessible and spatially close hotspots suffices and hotspots can be effectively used to steer the model.
    \item \textbf{Binder length:} The sequence length of the binder to be generated, measured in number of amino acid residues. 
    \item \textbf{Target cropping:} \themodel{} can be conditioned on cropped targets, for example in order to fit the context window. 
\end{enumerate}

The user should be aware of the following details on input representations:
\begin{itemize}
    \item \textbf{Context length:} The context length for inputs and outputs is 512 residues, counting jointly residues in the target and the binder.
    \item \textbf{Structurally unresolved residues:} Structurally unresolved residues in the target are dropped, which means that the amino acid identities of unresolved residues are not represented in the model's input either.
    \item \textbf{Target backbone only:} Only the backbone atoms of the target structure are provided to the model.
\end{itemize}

\subsection[Model inference speed]{Model inference speed}
\label{app:inference_speed}

\begin{figure}[h]
    \centering
    \includegraphics[width=0.5\linewidth]{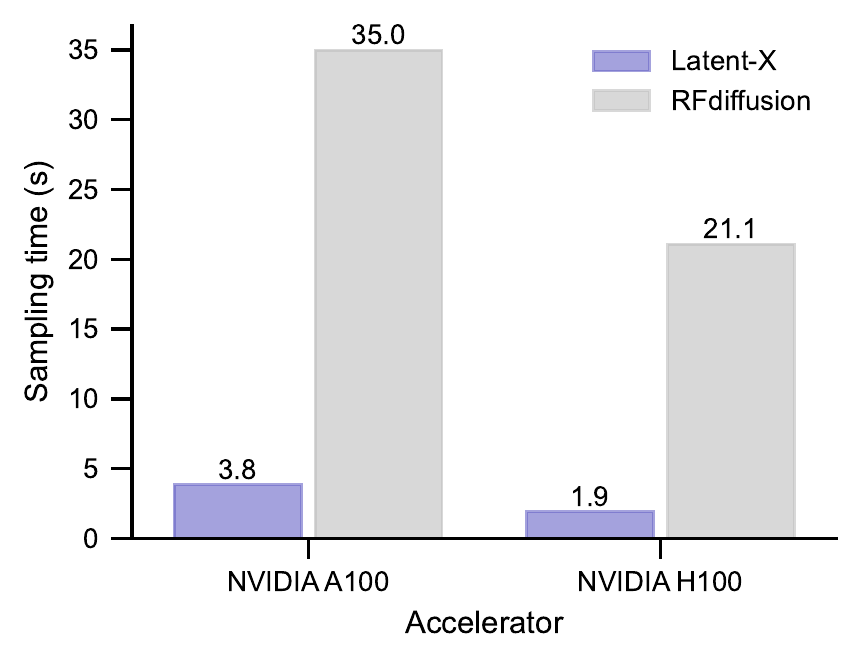}
    \customcaption{\textbf{Model inference speed on different GPU accelerators.} The sampling time corresponds to the time needed to sample a single \gls{pdl1} binder of 80 residues in seconds. Since \gls{pdl1} has 116 residues, this corresponds to a total of 196 generated residues. For a fair comparison, we did not use batching when sampling with \themodel{}.}
    \label{fig:inference_speed}
\end{figure}

We compared the inference speed between \themodel{} and \rfdiff{} on a simple benchmark task: generating a single binder of 80 amino acid residues for \gls{pdl1}. We recorded the runtime needed for this task on two common accelerators: NVIDIA A100 and NVIDIA H100 GPUs. We repeated this task 50 times and averaged runtime over the repetitions in order to control for potential variations between runs. The results are reported in \cref{fig:inference_speed}. 

We observe that \themodel{} is approximatively 10 times faster than \rfdiff{} for this task on both GPU accelerators. We attribute these speed gains to our proprietary architecture, which we optimized for fast inference. \themodel{} also benefits from modern GPU architectures, which is illustrated by a factor two speed-up when upgrading from a A100 to a H100 GPU.

Note that \themodel{} can be used to generate binders in batches. However, the standard \rfdiff{} workflow described in \cref{app:rfdiffusion} generates binders one by one. Hence, for a fair comparison, we did not use batching when sampling binders with \themodel{} in this runtime benchmark.

\subsection[Structural validity of samples]{Structural validity of samples}
\label{app:structural_validity}

We assessed the stereochemical quality of the all-atom structures for all binders selected for experimental validation by computing standard structural violations metrics as computed in \af{2} \cite{jumper2021highly}, see also \cite{mariani2013lddt}. In doing so we calculated peptide-bond geometry violations that measure the fraction of residues whose C–N bond length or backbone angles (CA–C–N or C–N–CA) deviate by more than 12 standard deviations (12$\sigma$) from the respective expected values from experimental structures. Inter-residue steric clashes give the fraction of residues that contain any atom overlapping another residue’s atom by more than \SI{1.5}{\angstrom} beyond the combined van-der-Waals radii. We also computed extreme \ca–\ca{} separation as the fraction of consecutive residue pairs whose \ca–\ca{} distance is more than \SI{1.5}{\angstrom} longer than the canonical \SI{3.8}{\angstrom}, suggesting potential backbone breaks. Further, we measure intra-residue clashes and internal geometry violations as the fraction of residues that exhibit self-clashes or distortions in bond-length or angle within the residue’s own atoms, using the same $12\sigma$ and \SI{1.5}{\angstrom} thresholds.

The analysis revealed high structural fidelity across all binders. Extreme \ca-\ca were completely absent in all our designs, with all binders for every target showing zero violation, indicating uninterrupted peptide backbones throughout the structures. Similarly, peptide bond geometry violations were non-existent.

Inter-residue steric clashes represented the only metric that deviated from the zero baseline though violations remained minimal overall. The majority of designs maintained clash rates of zero, while a minority of samples (\gls{il7ra}: 9\%, \gls{pdl1}: 1\%, \gls{sc2rbd}: 9\%, BHRF1: 23\%) observe a singular clash between two residues, which corresponds to a clash rate per residue of approximately 1\% or less. Intra-residue clashes and internal geometry problems were absent, remaining at zero across all targets and designs.

The overall per-residue violation rate therefore follows the inter-residue clash pattern, staying below 1\% for almost all structures and at zero for the majority of designs. Importantly, all observed violations could be resolved by applying the Amber relax protocol as applied in \af{2} \cite{jumper2021highly}, demonstrating the high stereochemical quality of \themodel{}'s protein binders.

\subsection[Target co-generation]{Target co-generation}
\label{app:target_sidechains}

\begin{figure}[h]
    \centering
    \includegraphics[width=\linewidth]{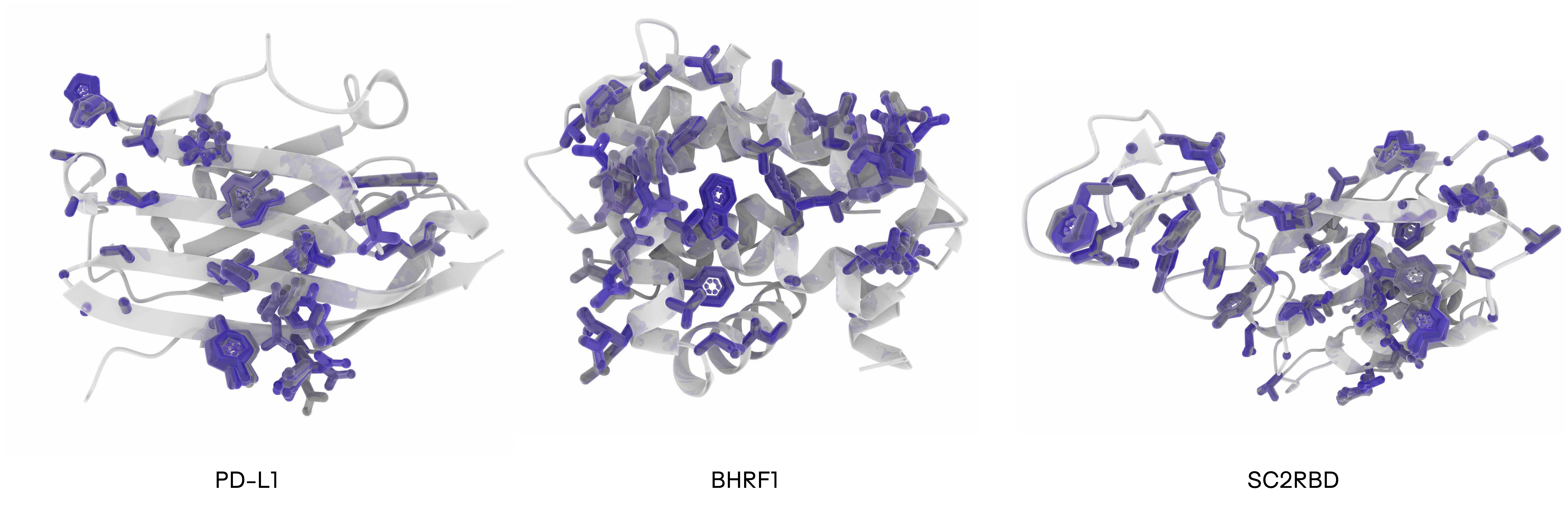}
    \customcaption{\textbf{Co-generation of target side chain rotamers.} The top view of three different targets: PD-L1, BHRF1, and SC2RBD, each with 10 superimposed structures from different binder design generations, selected at random. The side chains shown are the shared residues at the interface between the target and binder across the different designs. These side chains are coloured from grey to purple, with the gradient highlighting the variation in rotamers. The visible variation demonstrates the \themodel{}'s ability to adapt target rotamers in the target/binder interface.}
    \label{fig:target_sidechains}
\end{figure}

\subsection[Macrocycles with disulphide bonds]{Macrocycles with disulphide bonds}
\label{app:stapled_macrocycles}

\begin{figure}[h]
    \centering
    \includegraphics[width=\linewidth]{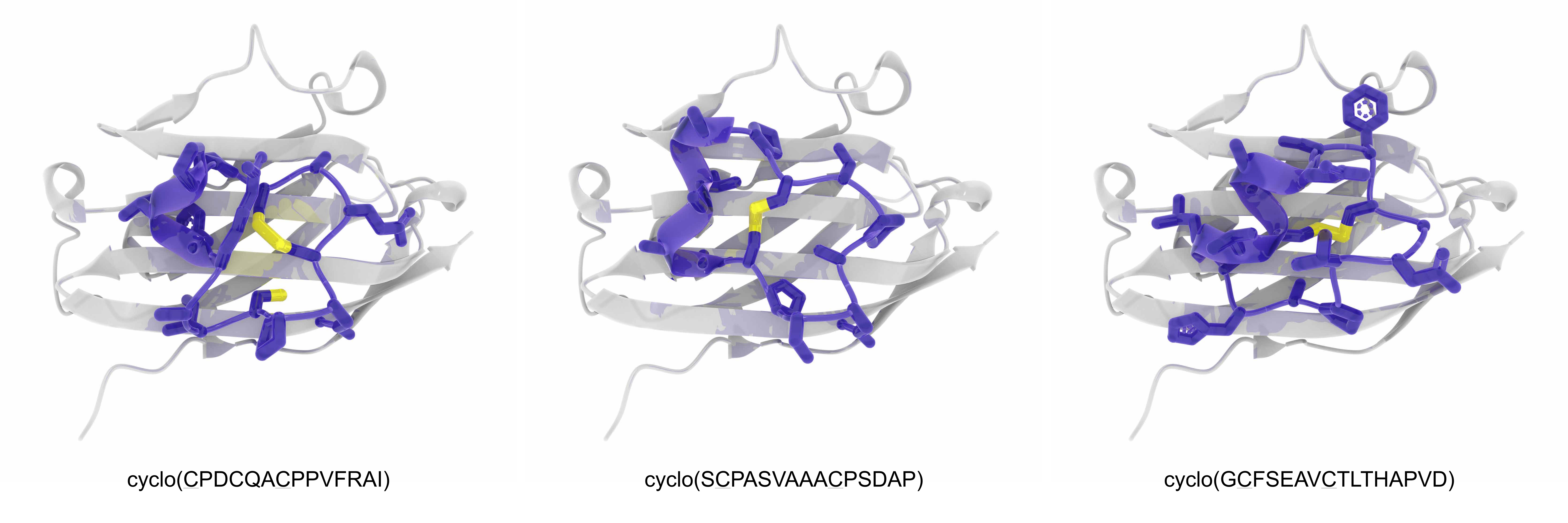}
    \customcaption{\textbf{Macrocycles with generated disulphide bonds.} Macrocycle designs generated against PD-L1, featuring intramolecular disulphide bridges formed between cysteine residues. Approximately 3\% of the 700 designs from the PD-L1 campaign contain more than one cysteine, allowing for disulphide bond formation. The designs displayed are randomly selected macrocycles with intramolecular disulphide bridges that passed the macrocycle \insilico{} filter. From left to right, the designs consist of 14, 15, and 16 amino acids. The sequences are shown, with cysteine residues underlined.}
    \label{fig:disulphide_macrocycle}
\end{figure}

\section[Tuning \insilico{} binding filters based on Structure Prediction Models]{Tuning \insilico{} binding filters based on structure prediction models}
\label{app:insilico_filter}

Computationally designed binders can be screened in order to increase the chance of experimental success using metrics derived from computational structure prediction as a proxy for binding success \cite{bennett2023improving, watson2023novo}. Following the procedure outlined in \cite{bennett2023improving, zambaldi2024novo}, we tuned an \insilico{} filter which we used to select designed binders for experimental testing. Due to commercial restrictions, we did not use \af{3} \cite{abramson2024accurate}. Instead, we tuned filters based on \chai{} \cite{chai2024chai} and \boltz{} \cite{passaro2025boltz}. These models reproduce the \af{3} architecture and approach \af{3} protein structure prediction performance. We used \chai{} based filtering in our experiments, but show in \cref{sec:expected_comp_hit_rate_boltz} that \chai{} and \boltz{} have qualitatively similar filter performance, demonstrating that the choice of latest generation structure prediction model is not critical for filtering. The \af{3} filter values quoted in this section are taken from the values reported in \cite{zambaldi2024novo}.

As input to the structure prediction models we provided sequences for both the binder and the target and provided the target structure as template. We did not use multiple sequence alignments and did not use templates for the binder. While \chai{} provides the option to use protein language model embeddings, we did not make use of this feature. For both models, we ran 10 trunk recycles and sampled 5 outputs from the diffusion head. We selected the highest confidence predicted structure from this set of structures according to \gls{iptm} and \gls{ptm} using the composite confidence metric $0.2 \ \gls{iptm} + 0.8 \ \gls{ptm}$, inline with \cite{zambaldi2024novo}. 

In order to determine retrospective ability to discriminate binders for all produced structure prediction metrics, we ran inference over the \num{640000} experimentally characterized \denovo{} designed mini-binder and target complexes from the Cao et al. dataset \cite{cao2022design}. This dataset contains binders across 11 targets, ranging from \num{15000}--\num{100000} binders per target. The dataset also provides a binding label per complex which was determined by yeast display. 

\begin{figure}[h]
    \centering
    \includegraphics[width=0.9\textwidth]{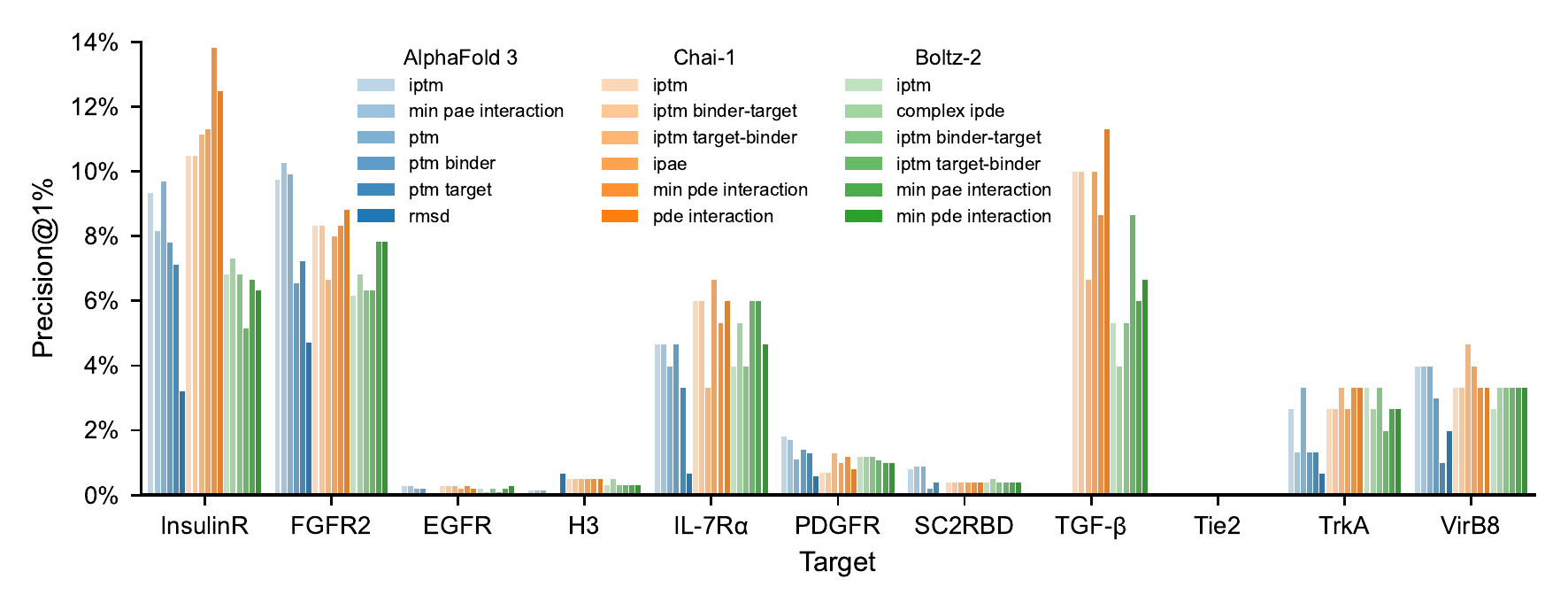}
    \customcaption{\textbf{Retrospective precision@1\% for \insilico{} metrics evaluated on the Cao et al. dataset.} The precision@1\% corresponds to the proportion of true binders in the top 1\% ranked binders, after retrospectively ranking the binders for each target by the respective structure prediction metric. The \insilico{} metrics for each structure prediction model are shown in average rank-order.}
    \label{fig:cao_metrics}
\end{figure}

After inference, we computed precision@1\% by computing precision in the top 1\% of samples according to six structure prediction metrics. \cref{fig:cao_metrics} shows the output metrics for \chai{} and \boltz{} selected by their average rank for precision@1\% across targets. We found the precision@1\% of \chai{} and \boltz{} to be comparable to \af{3} across most targets. The notable exception is TGF-$\beta$. This target consists of three short segments spanning two chains. A possible reason for the performance differences for this target may lie in how multiple chains were handled when constructing templates for \af{3} in \cite{zambaldi2024novo}.

\begin{figure}[h]
    \centering
    \includegraphics[width=0.9\textwidth]{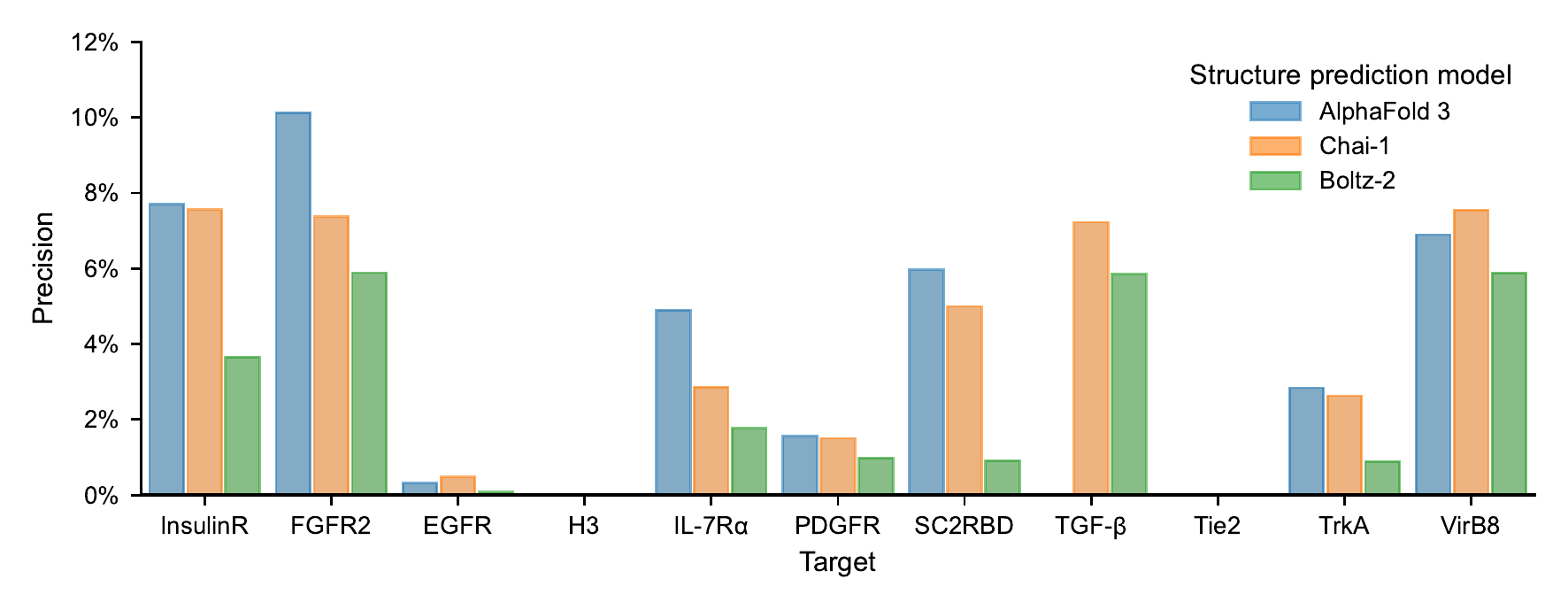}
    \customcaption{\textbf{Retrospective precision of the \insilico{} filters applied to the Cao et al. dataset.} We compared \insilico{} filters based on various structure prediction models. The precision is the fraction of true binders as identified by applying the \insilico{} filters using the thresholds given in \cref{tab:in_silico_filter_thresholds}.}
    \label{fig:cao_filter}
\end{figure}

We replicated the filter tuning procedure described in \cite{zambaldi2024novo} for \chai{} and \boltz{}. This procedure involved a grid search over metric thresholds, and selecting optimal thresholds based on the average rank over targets. Aligning with \cite{zambaldi2024novo}, the grid search included the \gls{minipae}, the \gls{ptmbinder}, and the \gls{complexrmsd}. We report the resulting optimal filter thresholds in \cref{tab:in_silico_filter_thresholds}. The precision of each filter retrospectively applied to the Cao et al. dataset is shown in \cref{fig:cao_filter}. The precision of the filter is target dependent but broadly comparable across structure prediction methods.

\gls{minipae} is the minium predicted aligned error in the binding interface and is a measure of confidence for the predicted relative placement of the binder and the target structure. \gls{ptmbinder} is the predicted TM-score \cite{xu2010significant} for the binder and is measure of confidence for the predicted binder structure. \gls{complexrmsd} is the root mean square deviation between the generated and the predicted structure of the complex formed by binder and target.

\begin{table}[h!]
\centering
\begin{tabular}{lccc}
Structure Prediction Model & \gls{minipae} & \gls{ptmbinder} & \gls{complexrmsd} \\
\midrule
AlphaFold 3     & $< 1.5$           & $> 0.8\phantom{0}$ & $< 2.5$           \\
Chai-1          & $< 1\phantom{.0}$ & $> 0.9\phantom{0}$ & $< 2\phantom{.0}$ \\
Boltz-2         & $< 1\phantom{.0}$ & $> 0.95$           & $< 2.5$           \\
\end{tabular}
\vspace{0.3cm}
\customtablecaption{\textbf{Comparison of \insilico{} filter thresholds.} Metric thresholds for each structure prediction model resulting from tuning on the Cao et al. mini-binder dataset.}
\label{tab:in_silico_filter_thresholds}
\end{table}

\subsection{Relaxed \insilico{} filter for macrocycles}
\label{app:insilico_filter_macrocycles}
We found the \insilico{} filters required modification for use with macrocyles. Using cyclic positional encodings in the structure prediction models, we predicted the structure for three macrocycles comprised of only canonical amino acids in complex with targets (PDB IDs: \texttt{9cdz}, \texttt{7oun}, \texttt{1sfi}). Despite being experimentally validated binders, the predicted structures did not pass the \insilico{} filter due to \gls{ptmbinder} being consistently well below the filter threshold, see \cref{tab:chai_cyclic_on_pdb_ids}. Therefore, we opted to drop this component in the filter when filtering macrocyle binders, and only filter based on the \gls{complexrmsd} and \gls{minipae} using the same thresholds given in \cref{tab:in_silico_filter_thresholds}. Note that this choice of filter for macrocycles was post-hoc justified by the high resulting experimental success, see \cref{sec:macro_results}.  %

\begin{table}[h!]
\centering
\begin{tabular}{lccc}
PDB ID & \gls{minipae} & \gls{ptmbinder} & \gls{complexrmsd} \\
\midrule
\texttt{9cdz}  & $0.49$ & $0.19$ & $0.50$ \\
\texttt{7oun}  & $0.38$ & $0.18$ & $0.73$ \\
\texttt{1sfi}  & $0.27$ & $0.22$ & $0.77$ \\
\end{tabular}
\vspace{0.3cm}
\customtablecaption{\textbf{Structure prediction metrics for exemplary macrocycle complexes in the \gls{pdb}.} The \gls{minipae} and \gls{complexrmsd} values meet the threshold criteria for the mini-binder \insilico{} filter, but the value of \gls{ptmbinder} is consistently well below the mini-binder passing threshold. The metrics shown were produced with \chai{}.}
\label{tab:chai_cyclic_on_pdb_ids}
\end{table}

\section[Computational design and filtering]{Computational design and filtering}
\label{app:computational_design}

\subsection{Macrocycles}
\label{app:computational_design_macrocycles}
This section gives additional details for the automated filters used to select  macrocycles for lab validation, see \cref{sec:comp_design_pipeline_macro}. 

To assess the sequence novelty of our designed macrocycles we implemented a sequence-based filtering pipeline against the \gls{pdb}. First, we made a peptide database by extracting all protein chains from the \gls{pdb} that were $\leq 30$ amino acids. For each designed macrocycle, we generated all possible cyclic permutations, where a 12-residue peptide yields 12 unique rotations, which were then used as queries. We performed sequence similarity searches using MMseqs2 \cite{steinegger2017mmseqs2} with parameters optimized for short peptide alignment. The exact command we used is as follows: \texttt{mmseqs search queryDB targetDB resultDB tmp -s 7.5 --num-iterations 3 --max-seqs 300 --spaced-kmer-pattern 110111 -k 5 --exact-kmer-matching 1 -e inf}. A sequence was considered a match and filtered out if any of its cyclic permutations exhibited greater than 50\% sequence identity. %

Macrocyclic peptides can be challenging to synthesize, and prior studies have reported high failure rates at this stage \cite{rettie2025accurate, rettie2025cyclic}. To mitigate this risk, we applied empirical synthesis filters based on vendor guidelines (GenScript). Sequences were excluded if they contained fewer than 40\% hydrophilic residues, consecutive identical residues, more than four consecutive hydrophobic residues or any cysteine residues.

\subsection{Mini-binders}
\label{app:computational_design_minibinders}
This section gives additional details for the automated filters used to select  mini-binders for lab validation, see \cref{sec:comp_design_pipeline_mini}. 

To evaluate the sequence novelty of generated mini-binders, we followed the protocol outlined in \cite{verkuil2022language}. We queried each sequence against UniRef50 \cite{suzek2015uniref} using MMseqs2 \cite{steinegger2017mmseqs2} with the following command: \texttt{mmseqs search queryDB uniref50 resultDB tmp -s 7.5 --num-iterations 3 -e 1 --max-seqs 300}. This search identified all matches with an E-value $\leq 1$. From these matches, we applied a sequence identity filter: sequences were considered novel if none of the identified matches had a percent identity above 20%

After filtering for novelty, we performed structural clustering using Foldseek \cite{van2024fast} with the following command: \texttt{foldseek easy-cluster structures result tmp --alignment-type 1 --tmscore-threshold 0.6}. We selected designs from across the resulting structural clusters to ensure our experimental validation covered a diverse range of generated structures.

\section[Methods for estimating expected \insilico{} hit rates on unseen targets
]{Methods for expected \insilico{} hit rates on unseen targets}
\label{app:expected_comp_hit_rate}

\subsection[Test set construction]{Test set construction}
\label{app:testset_construction}

We downloaded structures from the \gls{pdb} released after 24 November 2023. This date is beyond the cut-off date for the training datasets of Latent-X, \rfdiff{}, \rfpep{}, ProteinMPNN, \chai{} and \boltz{}. To ensure high quality target structures, we filtered out NMR structures, applied a resolution filter (<\SI{3}{\angstrom}), excluded assemblies containing DNA or RNA, and selected structures such that all chains are over 90\% structurally resolved. Furthermore, we excluded C2 and D2 symmetries, the most common homodimer symmetry groups, and selected bioassemblies with 1 to 5 chains. From each of the bioassemblies, we randomly selected an individual chain in the length range 30--400 to serve as a target.

To promote diversity in the target proteins, we downloaded sequence clusters at a threshold of 40\% sequence homology from the \gls{pdb}. After removing singleton clusters, we randomly sampled a single representative from each cluster. We used a hotspot selection algorithm to automatically select three non-overlapping epitopes per target, see \cref{app:hotspot_picker} for more detail. For some targets we were unable to select three unique epitopes, for example because the protein did not have sufficient accessible surface area, and these were discarded.

One of the three metrics comprising the \insilico{} filter is \gls{complexrmsd}. Despite the target template being provided to the structure prediction model, we observed a small number of targets structures being predicted incorrectly irrespective of the binder quality, leading to large \gls{complexrmsd}. We observed two scenarios where this could occur: the first is target proteins with elongated folds where a ``hinge effect" can lead to large overall \gls{complexrmsd}. The second scenario is loopy terminal regions in the target protein, leading to poor agreement between the predicted and true target structure. To prevent these problematic cases, and arrive at the final 200 structures, we therefore excluded elongated target folds based on radius of gyration and excluded targets with low structural agreement by determining the structural agreement of predicted structures of the target without binder. For the latter we used no multiple sequence alignment and only the ground truth structure as template, as it was done for the \insilico{} filter. To arrive at the final 200 target structures we selected the remaining targets with the lowest maximum \gls{rmsd} of the target structure from 50 diffusion head samples. 

\subsection[Automated hotspot selection for \insilico{} studies]{Automated hotspot selection for \insilico{} studies}
\label{app:hotspot_picker}

To assess \insilico{} binding hit rate on unseen targets, we used an automated algorithm to select plausible binding hotspots, described in the following. Candidate residues were required to be surface-exposed and located in ordered secondary structure. To this end, we computed relative accessible surface area (rASA) using Biopython \cite{cock2009biopython} and assigned secondary structure using DSSP \cite{kabsch1983dictionary}. Residues were retained if they have rASA > 30\% and are classified as 'H' (helix), 'E' (sheet), 'G' ($3_{10}$ helix), or 'B' (beta-bridge). Each epitope was sampled to consist of three residues within \SI{10}{\angstrom} of each other, sampled from qualifying candidate residues. When testing multiple epitopes on the same targets, we required the epitopes to be non-overlapping: no residue can be reused, and residues within an epitope must be spaced by at least one position in the primary sequence (i.e., no adjacent residues). Note that residues on loops were excluded in the automated hotspot picking. This is a conservative choice that we find does not matter in practice.

\subsection{Expected \insilico{} hit rates on unseen targets with an alternative structure prediction model}
\label{sec:expected_comp_hit_rate_boltz}

In addition to using an \insilico{} filter based on \chai{} as a structure prediction model to validate our designs and for computing prospective hit rates for \rfdiff{}-based and \themodel{} workflows, we also scored all generated binders with the \insilico{} filter we tuned based on \boltz{}, as detailed in \cref{app:insilico_filter}. The results are depicted in \subfigref{fig:dist_boltz}{a} for macrocycles and in \subfigref{fig:dist_boltz}{b} for mini-binders.

As with scoring based on the \chai{} \insilico{} filter, \themodel{} compares favourably to \rfpep{} with an average success rate per target of 5.33\% compared to 1.19\% for the \rfdiff{}-based workflow for macrocycles.
For mini-binders as well, \themodel{} outperforms \rfdiff{} with an average success rate per target of 3.92\% compared to 1.93\%.

\begin{figure}[h!]
    \centering
    \includegraphics[width=0.9\linewidth]{figures/S6_distplot_boltz.pdf}
    \customcaption{\textbf{\textit{In silico} hit rates for \themodel{} vs \rfpep{} / \rfdiff{} using an \insilico{} filter based on an alternative structure prediction model.} \textbf{a)} Distribution of \insilico{} hit rates for macrocycles on 200 held-out targets. \themodel{}'s hit rates are shown in orange, and those of \rfpep{} in grey. \textbf{b)} Distribution of \insilico{} hit rates for mini-binders on 200 held-out targets. \themodel{}'s hit rates are shown in purple, and those of \rfdiff{} in grey. Distributions show the binned relative frequency against the $y$-axis shown on the left. Curves show the cumulative distribution functions against the $y$-axis on the right. Dashed vertical lines show the \insilico{} success rates achieved in design workflows for lab validation, see \cref{sec:comp_design_pipeline_macro} and \cref{sec:comp_design_pipeline_mini}. \textit{In silico} filter based on \boltz{}.}
    \label{fig:dist_boltz}
\end{figure}

\subsection[\rfdiff{} and \rfpep{} workflow]{\rfdiff{} and \rfpep{} workflows}
\label{app:rfdiffusion}

To generate binders with \rfdiff{} or \rfpep{} we followed prior work \cite{rettie2025cyclic, watson2023novo, zambaldi2024novo} and downloaded \rfdiff{} from the official RosettaCommons \rfdiff{} GitHub repository.\footnote{https://github.com/RosettaCommons/RFdiffusion}

\paragraph{Macrocycles}
For macrocycles, we used the \rfpep{} pipeline \cite{rettie2025cyclic} which corresponds to the \rfdiff{} workflow with additional cyclic constraints applied to the \rfdiff{} model and a reduced number of 50 diffusion steps for generation. Sequence generation was then carried out using ProteinMPNN using the same settings as for mini-binders, see below.

All designs used hotspot definitions and binder length parameters identical to our own pipeline, see \cref{sec:benchmarking_protocol}, to ensure fair comparative analysis. 

\paragraph{Mini-binders}
We generate mini-binders with the \texttt{noise=0} setting which has been shown to produce higher \insilico{} binding success \cite{watson2023novo} than higher noise levels. 

As is required and recommned in the \rfdiff{} workfow, we used ProteinMPNN \cite{dauparas2022robust} to generate eight sequences for each \rfdiff{} generated binder backbone using the default low-temperature setting of $10^{-4}$ and selected the sequence with the highest average residue log-probability over the generated binder.

\subsection[Structural diversity of successful mini-binder designs]{Structural diversity of successful mini-binder designs}

\begin{figure}[h]
    \centering
    \includegraphics[width=0.5
    \textwidth]{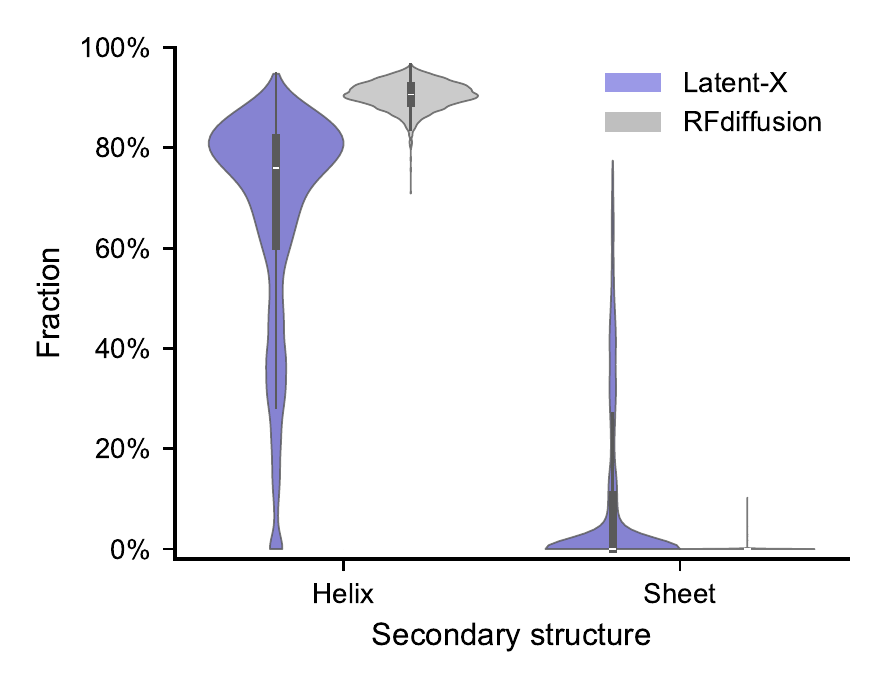}
    \customcaption{\textbf{Secondary structure fractions for mini-binders that pass our \insilico{}-filter.} Fraction of residues annotated as $\alpha$-helix or strand / sheet for binders generated by \themodel{} (purple) and \rfdiff{} (grey) on the test set of 200 targets, used in \cref{sec:expected_comp_hit_rate}, that pass the \insilico{} filter.}
    \label{fig:dist_plot_structural_diversity}
\end{figure}

We calculated the diversity of secondary structures for mini-binders passing the \insilico{} filter based on \chai{}, on the test set of 200 PDB structures, used in \cref{sec:expected_comp_hit_rate}. We used the DSSP algorithm \cite{kabsch1983dictionary} to determine secondary structure fractions considering both $\alpha$-helices and $\beta$-sheets.
As shown in \cref{fig:dist_plot_structural_diversity} \themodel{} produced a much more diverse distribution of secondary structures than \rfdiff{}, which mostly generated helical bundles. In contrast, \themodel{} produced more diverse folds with varying secondary structure, including structures with considerable fractions of $\beta$-sheets which are mostly absent in \rfdiff{} generated binders.

\begin{table}[htb]
    \centering
    \begin{tabular}{l>{\centering\arraybackslash}p{1.6cm}>{\ttfamily\centering\arraybackslash}p{0.7cm}>{\ttfamily\arraybackslash}p{1.6cm}>{\ttfamily\arraybackslash}p{2.6cm}>{\centering\arraybackslash}p{2.5cm}>{\ttfamily\arraybackslash}p{1.0cm}}
    \raisebox{-1ex}{Target} & Binder modality & \textrm{PDB ID} & \centering\textrm{Target chain and residues} & \centering\textrm{Target hotspot residues} & Natural binding partner & \textrm{Binder length} \\
    \midrule
    MDM2 & Macrocycle & \seqsplit{4hfz} & \seqsplit{A26-108} & {A54,A58,A61} & p53 helix & \seqsplit{12-18} \\
    MCL-1 & Macrocycle & \seqsplit{2pqk} & \seqsplit{A172-197,A203-321} & A224,A227,A231,A235, A249,A253,A263 & BH3 helix & \seqsplit{12-18} \\
    PD-L1 & Macrocycle & \seqsplit{5o45} & \seqsplit{A17-132} & A56,A115,A123 & PD-1 & \seqsplit{12-18} \\
    \\[1em]
    BHRF1 & Mini-binder & \seqsplit{2wh6} & \seqsplit{A2-158} & A65,A74,A77,A82, A85,A93 & BH3 helix & \seqsplit{80-120} \\
    \gls{il7ra} & Mini-binder & \seqsplit{3di3} & \seqsplit{B17-209} & B58,B80,B139 & IL-7 & \seqsplit{80-120} \\
    PD-L1 & Mini-binder & \seqsplit{5o45} & \seqsplit{A17-132} & A56,A115,A123 & PD-1 & \seqsplit{80-120} \\
    SC2RBD & Mini-binder & \seqsplit{6m0j} & \seqsplit{E333-526} & E485,E489,E494, E500,E505 & ACE2 receptor & \seqsplit{80-120} \\
    TrkA & Mini-binder & \seqsplit{1www} & \seqsplit{X282-382} & X294,X296,X333 & Nerve growth factor & \seqsplit{80-120} \\
    [0.5em]
    \end{tabular}
    \vspace{0.3cm}
    \customtablecaption{\textbf{Benchmark protein targets used in experimental validation.} Summary of structural targets used, including \gls{pdb} identifiers, target regions, hotspot residues, natural binding partners, and binder length ranges used for experimental validation studies.}
\label{tab:target_table}
\end{table}

\begin{figure}[h]
    \centering
    \includegraphics[width=\textwidth]{figures/S8_all_macrocycles.pdf}
    \customcaption{\textbf{All successfully lab validated \themodel{} generated macrocycles.} We show 25 \denovo{} designed bound structures across 3 different targets with their corresponding binding affinities ($K_D$) determined via \gls{SPR}. The all-atom structure of \themodel{} generated macrocycles are shown in orange, while the targets are shown in grey, limiting to representations of their backbones for visual clarity.}
    \label{fig:macrocycle_wall_of_best_hits}
\end{figure}

\begin{figure}[h]
    \centering
    \includegraphics[width=\textwidth]{figures/S9_all_minibinders.pdf}
    \customcaption{\textbf{All successfully lab validated \themodel{} generated mini-binders.} We show 87 \denovo{} designed bound structures across 5 different targets with their corresponding binding affinities ($K_D$) determined via 5 concentration \gls{BLI}. The all-atom structure of \themodel{} generated mini-binders are shown in purple, while the targets are shown in grey, limiting to representations of their backbones for visual clarity.}
    \label{fig:minibinder_wall_of_best_hits}
\end{figure}

\clearpage

\begin{figure}[ht]
    \centering
    \includegraphics[width=0.9\linewidth]{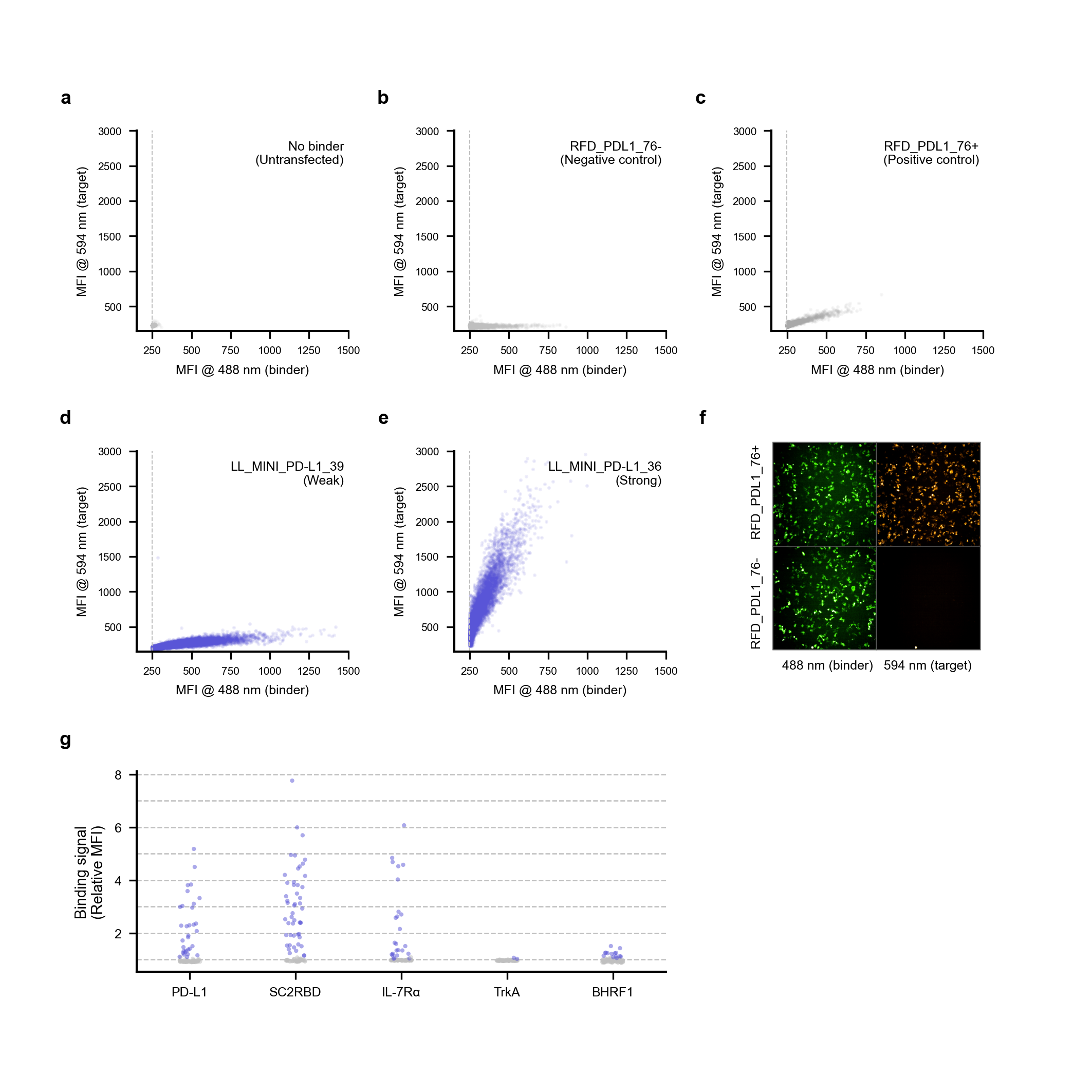}
    \vspace{-1cm}
    \customcaption{\textbf{High-throughput \gls{mDisplay} screening of computationally designed mini-binders.} 
    Each scatter plot represents one well within \gls{mDisplay} screening. To verify binder expression, the minimum threshold was determined by comparison with \textbf{a)} untransfected control; specifically, the 95th percentile of 488 nm fluorescence intensity in untransfected wells is indicated by a grey dotted line. RFD\_PDL1\_76 \cite{watson2023novo} served as an additional control, shown in \textbf{b)} without exposure to target protein PD-L1 and \textbf{c)} with exposure to PD-L1.
    \textbf{(d, e)} MFI measured for two \themodel{} designed binders, exemplifying the dynamic signal range for varying binding strengths. \textbf{f)} Confocal images within two wells of \gls{mDisplay} experiment. \textbf{g)} Binding signal for all 88 designs per target that were screened in \gls{mDisplay}, see \cref{sec:mini_results}. Purple points represent \denovo{} binders that exhibited binding signal significantly different from negative control (\cref{app:mdisplay}).}
    \label{fig:Definition of mDisplay}
\end{figure}

\begin{figure}
    \centering
    \includegraphics[width=1.0\linewidth]{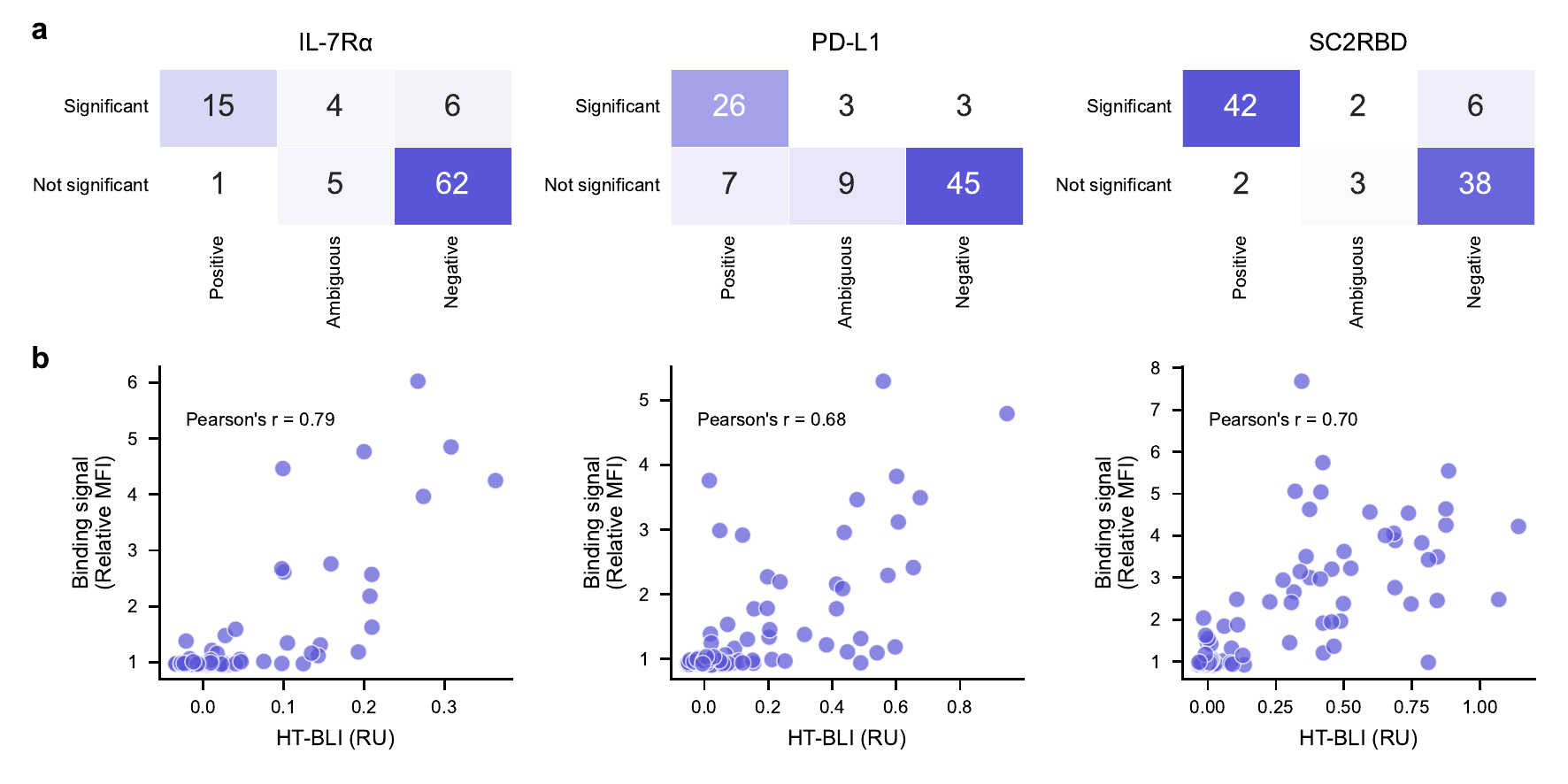}
    \customcaption{ \textbf{Correlation between \gls{mDisplay} and HT-BLI across three targets.} \textbf{a)} Confusion matrices comparing \gls{mDisplay} and HT-BLI results. Rows indicate \gls{mDisplay} classifications (“Significant” or “Not significant” as described in \cref{app:mdisplay}), while columns indicate categorical outcomes from HT-BLI. \textbf{b)} Scatter plots showing quantitative correlation between the two assays. The x-axis denotes HT-BLI response units (RU), and the y-axis indicates relative mean fluorescence intensity (MFI) measured by \gls{mDisplay}. Note, the three targets with the highest dynamic range MFI values in \gls{mDisplay} are shown, as seen in \cref{fig:Definition of mDisplay}.}
    \label{fig:mammalian HT-BLI correlation}
\end{figure}

\begin{figure}
    \centering
    \includegraphics[width=1.0\linewidth]{figures/S12_sensogram.pdf}
    \customcaption{\textbf{Raw sensogram data for macrocycles.} Raw sensogram data for macrocycles from an eight-concentration SPR experiment, with highest concentrations ranging from 25 to 500 µM. The \kd{} is stated on the plot.}
    \label{fig:raw_sensogram_macrocycle}
\end{figure}

\clearpage

\clearpage
\thispagestyle{empty}
\begin{sidewaystable}[p]
\centering

\begin{tabular}{@{}>{\fontsize{7}{8.4}\selectfont}l@{\hspace{2em}}>{\fontsize{7}{8.4}\selectfont}l@{\hspace{2em}}>{\fontsize{7.5}{8.4}\selectfont\raggedright\ttfamily\arraybackslash}p{16.4cm}@{}}
\textbf{\normalsize Binder} & \textbf{\normalsize \kd{} (M)} & \textbf{\normalsize\textrm{Amino acid sequence}} \\ 
\midrule
LL\_CYC\_PD-L1\_24 & $7.17\times 10^{-5}$ & \seqsplit{cyclo(SPTAKALFPSLSIPNQ)} \\
LL\_CYC\_PD-L1\_9 & $7.24\times10^{-5}$ & \seqsplit{cyclo(QERSNPAVPEVFRPILPE)} \\
LL\_CYC\_PD-L1\_28 & $1.17\times10^{-4}$ & \seqsplit{cyclo(IASKHGISTAGTHPEILK)} \\
LL\_CYC\_MDM2\_18 & $5.35 \times 10^{-6}$ & \seqsplit{cyclo(AEPVQADTFADYWRSL)} \\
LL\_CYC\_MDM2\_13 & $1.64 \times 10^{-5}$ & \seqsplit{cyclo(PAKLHLILAGEQSEFV)} \\
LL\_CYC\_MDM2\_6 & $1.71\times 10^{-5}$ & \seqsplit{cyclo(WLKLMGEDVPEDSFANY)} \\
RFP\_MDM2\_8 & $8.38\times 10^{-6}$ & \textrm{See \cite{rettie2025accurate}} \\
LL\_CYC\_MCL-1\_2 & $1.84 \times 10^{-5}$ & \seqsplit{cyclo(GSVTGPVADLMRAFGLDP)} \\
LL\_CYC\_MCL-1\_29 & $1.97 \times 10^{-5}$ & \seqsplit{cyclo(FGMDPGSVSGPVADLMRA)} \\
LL\_CYC\_MCL-1\_8 & $2.71 \times 10^{-5}$ & \seqsplit{cyclo(GPVADLMRAFGLDPGSIE)} \\
RFP\_MCL-1\_2 & $1.0 \times 10^{-5}$ & \textrm{See \cite{rettie2025accurate}} \\
LL\_MINI\_BHRF1\_58 & $2.25\times 10^{-8}$ & \seqsplit{STADTLLAGLAEAATLVAQGKIDEAVALLASLTPLVPALPRGAFGAVAARIAAVQILAIGLYEEGAITAAQRDAVVAAAKALGDALDAAAA} \\
LL\_MINI\_BHRF1\_13 & $2.34 \times 10^{-8}$ & \seqsplit{VTPAQVAKETAIILQLIGALEDLRLGRLDEAEAALAATQPRLGELSDAALLAHAAALLDAIEAEATAAGAAGIVALVGEMKAALAAV} \\
LL\_MINI\_BHRF1\_89 & $2.65\times 10^{-8}$ & \seqsplit{SVADQIAALAVQVEALSEKTSPAVLPTLSNAELEALLAEANSLLDQLAALSSTNLDDEIAIGILMIDLATAKVAIEGELKRRG} \\
BINDI  & $6.86\times 10^{-9}$ & \textrm{See \cite{procko2014computationally}} \\
GDM\_BHRF1\_70 & $4.58\times 10^{-7}$ & \textrm{See \cite{zambaldi2024novo}} \\
LL\_MINI\_IL-7Ra\_39 & $<1.00\times 10^{-11}$ & \seqsplit{SASSLQKDINALETDINTLETDTLTTLNTYNEKASSLDNDTKNKINDEFNTIRDNYINANTKLIEAKNLLNEGKVNEAQTKLNESKAALNTAKTALNNAKNLIAS} \\
LL\_MINI\_IL-7Ra\_69 & $2.21\times 10^{-9}$ & \seqsplit{SAAAQINTLKNASNLINEGKYGEAVTLLTNLLPTVADNAVKTEINNAIDNAKEADLYAKKADLAEQYLAENPGLSAADVDKVKYLIDAYKDISADSATAAKTSINNAISLLETNS} \\
LL\_MINI\_IL-7Ra\_58 & $1.63\times 10^{-8}$ & \seqsplit{ATYSKQITIPVAASDRAAVETVLNGKEFADGTIVPGIIARAKEAANLNVNYKYDGNNLIVTVTGSNEAAVNNTADNIKAILEG} \\
GDM\_IL-7RA\_70 & $<1.00\times 10^{-11}$ & \textrm{See \cite{zambaldi2024novo}} \\
RFD\_IL-7RA\_55 & $3.11 \times 10 ^{-8}$ & \textrm{See \cite{watson2023novo}} \\
LL\_MINI\_PD-L1\_34 & $2.69\times 10^{-10}$ & \seqsplit{GAELDKVKEELDAILKKVDDLTDAKFSSLTTAQAELYFKYKAEGNKIKEEADNLLAQGKTDEAIAKYKEAIEKYKQAAKVLENA}\\
LL\_MINI\_PD-L1\_36 & $4.71\times 10^{-10}$ & \seqsplit{SAATALKNEGIDLAKQADDKVKIAKKLKGEGKSKEAIAKYKEAVALYKDAATKFGQASKLFYEQEDFYESGLMAKEVVNMLKKAADMDDEIEKLEK} \\
LL\_MINI\_PD-L1\_10 & $3.71\times10^{-9}$ & \seqsplit{ADEIVSKFKEAIDNINEGIKFAKEGDAVKFKEYMKKAVESLKEAATLSVQAKKPEVALLALKAIGKASYAVDLADEKKFPEAVAVAKEAIELAKQGIEKAG} \\
GDM\_PD-L1\_135 & $<1.00\times 10^{-11}$ & \textrm{See \cite{zambaldi2024novo}} \\
RFD\_PD-L1\_76 & \nobinder & \textrm{See \cite{watson2023novo}} \\
LL\_MINI\_SC2RBD\_8 & $<1.00\times 10^{-11}$ & \seqsplit{ASFAALQATANAKLDELEAAVATAEAELAKASPNQATVQANLTKGNQLKMELADTIYDMQKSLEASPSAANTTAFGNVVSRYTSLTKRLVAAGQKYASLFG}\\
LL\_MINI\_SC2RBD\_35 & $<1.00\times 10^{-11}$ & \seqsplit{SAVIEGIASNYIGNKLKYSDDQLKEEIKFINDNVDDLNKKYFVNSVKNNKYLKDEVKDFSEDALTALGFYVSSFISFVQHEAKKNKEKADNDVINGLLEKYSDSFVELGKRFENVS}\\
LL\_MINI\_SC2RBD\_62 & $1.18\times 10^{-9}$ & \seqsplit{TYTYYPSAVGTVAPNSDGSVNLSFTTLASSSKGKAVAALSNGQTVKLKYTDKSGNTVTLDATLVGKVSDTVNGTSVIKYGLQISSANASEVTANLSDGSAIEVGIPN} \\
GDM\_SC2RBD\_104 & $1.67\times 10^{-10}$ & \textrm{See \cite{zambaldi2024novo}} \\
LCB1 & $1.88\times 10^{-10}$ & \textrm{See \cite{cao2020novo}} \\
LL\_MINI\_TrkA\_86 & $4.01\times 10^{-11}$ & \seqsplit{GAAAAAAAQDAANQNTANVVAKIAAQQTAEEVQNLAAAGQAIPSSLSYSSGDYTINVTIAQTGTNQYTITATVTYKGKTATATKTITLA}\\
LL\_MINI\_TrkA\_52 & $1.21\times 10^{-8}$ & \seqsplit{SLAPQTVNGTIKTSDSLTASQASGVGDFLARKTLSDAGKAIKSGDSSVSKVQYSLDKPISLSEVSAGKYEGSVEGKLIVTKKDGSVEEIPVVTNVTVQKTATGYKVSATTKAK}\\
LL\_MINI\_TrkA\_40 & $1.67\times 10^{-8}$ & \seqsplit{GLSSSETSSATSSLTSHLNSALSSKGLSLPASSKTAIVSDAMTIVSNSVPKLEAAGLPRDIAVKYAVASAKSYIDDVVSTIAA} \\
GDM\_TrkA\_9 & \nobinder & \textrm{See \cite{zambaldi2024novo}} \\
RFD\_TrkA\_88  & \nobinder & \textrm{See \cite{watson2023novo}} \\
\end{tabular}
\vspace{0.3cm}
\customtablecaption{\textbf{Binder sequences and their dissociation constants (\kd{}).} Selected binders including our two strongest reported binders and one additional representative binder per target. \kd{} values for mini-binders were obtained via \gls{BLI} whilst \kd{} values for macrocyles were obtained via \gls{SPR}. Reported \kd{} values for controls are replicate values. A cross (\nobinder) indicates no discernable binding was observed. Macrocycle sequences are denoted by \texttt{cyclo($\cdot$)}.}
\label{tab:Binder_sequences}
\end{sidewaystable}
\clearpage

\section{Experimental methods}
\label{sec: Experimental methods}
\subsection{Benchmarking against reference binders}
\label{app:benchmark}

To benchmark the performance of our mini-binders, we included the best previously published binders for BHRF1 \cite{procko2014computationally} and \gls{sc2rbd} \cite{cao2020novo}, and the best \rfdiff{}{} designs for \gls{il7ra}, \gls{pdl1}, and \gls{trka} \cite{watson2023novo} as positive controls in our \gls{mDisplay} and \gls{BLI} measurements seen in \cref{tab:positive-binders}. For \gls{BLI}, we also included the best binders for each target from \ap{} that represent the previous best-in-class designs for their respective targets \cite{zambaldi2024novo}. 
To benchmark the performance of our macrocycles, we included the best reported binder from \rfpep{} for \gls{mdm2} and \gls{mcl1} \cite{rettie2025accurate}.

As described in \cref{tab:binding_affinities}, there are observed differences among published and replicated data, as is within expectation, likely due to differences in materials and assay protocols. RFD\_PDL1\_76, RFD\_TrkA\_88, and GDM\_TrkA\_9 all failed to produce measurable signal in both \gls{HT-BLI} and 5-concentration \gls{BLI}. However, all five reference mini-binders yielded detectable signal in \gls{mDisplay}, including RFD\_PDL1\_76 and RFD\_TrkA\_88. Out of the three binders that did not work, only RFD\_PDL1\_76 yielded a \kd{} upon follow-up with SPR.

\begin{table}[htbp]
\centering
\scalebox{0.95}{
\begin{tabular}{@{}>{\arraybackslash}p{1.05cm}>{\arraybackslash}p{2.6cm}lr|l>{\centering\arraybackslash}p{1.3cm}>{\centering\arraybackslash}p{0.4cm}>{\centering\arraybackslash}p{0.4cm}c}
\textbf{Target} & \textbf{Name} & \textbf{Type} & 
\multicolumn{2}{c}{\textbf{K\textsubscript{D} (nM)}} &
\textbf{mDisplay} & \textbf{BLI} & \textbf{SPR} & \textbf{Ref.} \\
\cline{4-5}
& & & \textbf{Published} & \textbf{Replicated} & & & & \\
\midrule
\gls{il7ra} & RFD\_IL7RA\_55 & Mini-binder & 30 & 31.1 & \checkmark & \checkmark & -- & \cite{watson2023novo} \\
\gls{il7ra} & GDM\_IL-7RA\_70 & Mini-binder & 0.08 & $<0.01$ & -- & \checkmark & -- & \cite{zambaldi2024novo} \\
\gls{pdl1} & RFD\_PDL1\_76 & Mini-binder & 1.6 & 699 & \checkmark & \checkmark & \checkmark & \cite{watson2023novo, zambaldi2024novo} \\
\gls{pdl1} & GDM\_PD-L1\_135 & Mini-binder & 0.18 & $<0.01$ & -- & \checkmark & -- & \cite{zambaldi2024novo} \\
\gls{sc2rbd} & LCB1 & Mini-binder & $<1$ & 0.188 & \checkmark & \checkmark & -- & \cite{cao2020novo} \\
\gls{sc2rbd} & GDM\_SC2RBD\_104 & Mini-binder & 26 & 0.167 & -- & \checkmark & -- & \cite{zambaldi2024novo} \\
\gls{trka} & RFD\_TrkA\_88 & Mini-binder & 328 & \nobinder & \checkmark & \checkmark & \checkmark & \cite{watson2023novo} \\
\gls{trka} & GDM\_TrkA\_9 & Mini-binder & 0.96 & \nobinder & -- & \checkmark & \checkmark & \cite{zambaldi2024novo} \\
BHRF1 & BINDI & Mini-binder & $0.22 \pm 0.05$ & $6.86$ & \checkmark & \checkmark & -- & \cite{procko2014computationally} \\
BHRF1 & GDM\_BHRF1\_70 & Mini-binder & 8.5 & 458 & -- & \checkmark & -- & \cite{zambaldi2024novo} \\
\gls{mdm2} & RFP\_MDM2\_8 & Macrocycle & \num{1900} & \num{8380} & -- & -- & \checkmark & \cite{rettie2025cyclic} \\
\gls{mcl1} & RFP\_MCL-1\_2 & Macrocycle & \num{2000} & \num{10000} & -- & -- & \checkmark & \cite{rettie2025cyclic} \\
\end{tabular}
}
\vspace{0.3cm}
\customtablecaption{\textbf{Mini-binders and macrocycles used as positive controls and reference binders in experiments.} Published K\textsubscript{D} values are cited; replicated values shown where available. Checkmarks indicate which binding experiments the relevant binder was used in.}
\label{tab:positive-binders}
\end{table}

\subsection{Protein expression and purification}
\label{app:Protein expression and purification}
Mini-binder proteins were produced by GenScript Biotechnology Co., Ltd. (Nanjing, China), referred to as ‘‘GenScript’’ in this paper. Mini-binder DNA sequences were codon-optimized and synthesized by GenScript. The synthesized gene was subcloned into the pIVEX vector with an N-terminal his-tag for protein expression in a cell-free system. For \gls{HT-BLI}, cell-free protein synthesis reactions were assembled by combining S30 cell lysate, synthesis buffer, required enzymes, and plasmid DNA in a 48-deep-well plate. Each reaction had a final volume of \SI{1}{mL} and was incubated at \SI{25}{\celsius} for \SI{16}{hours}. The reaction mixture was then centrifuged at \SI{4000}{rpm} for \SI{5}{minutes}, and the supernatant was collected for analysis by SDS-PAGE. Samples were mixed with 5× reducing loading buffer (\SI{300}{mM} Tris-HCl pH 6.8, \SI{10}{\percent} SDS, \SI{30}{\percent} glycerol, \SI{0.5}{\percent} bromophenol blue, and \SI{250}{mM} DTT). Proteins were separated on a 4–\SI{20}{\percent} gradient SDS-PAGE gel (GenScript, cat. no. M42012) and visualized accordingly. For 5-concentration BLI, cell-free protein synthesis reactions were assembled by combining S30 cell lysate, synthesis buffer, required enzymes, and plasmid DNA in a 24-deep-well plate. Each reaction had a final volume of \SI{5}{mL} and was incubated at \SI{30}{\celsius} for \SI{6}{hours}. The reaction mixture was then centrifuged at \SI{4000}{rpm} for \SI{5}{minutes}, and the supernatant was collected for purification. Target protein was obtained by Ni column purification using Hamilton instruments. Target protein was sterilized by \SI{0.22}{\micro\meter} filter before stored in aliquots. The concentration was determined by A280 protein assay with BSA as standard. Protein purity was determined by standard SDS-PAGE confirmation. Target protein was analysed by SDS-PAGE. Samples were mixed with 5x reducing loading buffer (\SI{300}{mM} Tris-HCl pH 6.8, \SI{10}{\percent} SDS, \SI{30}{\percent} glycerol, \SI{0.5}{\percent} bromophenol blue, and \SI{250}{mM} DTT). Proteins were separated on a \SI{12}{\percent} homogeneous SDS-PAGE gel (GenScript, cat. no. M00668) and visualized accordingly. Purified proteins were verified to have >\SI{70}{\percent} purity before proceeding to 5-concentration \gls{BLI}.

BHRF1 (UniProt P03182, 2-160) with N-terminal his-tag, \gls{mdm2} (UniProt Q00987, 1-188), and \gls{mcl1} (UniProt Q07820, 172-327)  proteins were produced by GenScript as follows. The DNA sequences were codon-optimized for \textit{E. coli}, synthesized, and cloned into the pET30a vector with strep-tag (BHRF1) or his-tag (\gls{mdm2} and \gls{mcl1}). To evaluate expression, \textit{E. coli} strain BL21 Star™ (DE3) was transformed with recombinant plasmid. A single colony was inoculated into LB medium containing related antibiotic, culture was incubated in \SI{37}{\celsius} at \SI{200}{RPM} and then induced with IPTG. SDS-PAGE was used to monitor the expression. For expression, transformed BL21 Star™ (DE3) stored in glycerol was inoculated into TB medium containing related antibiotic and cultured at \SI{37}{\celsius}. When the OD600 reached about 1.2, cell culture was induced with IPTG at \SI{15}{\celsius}/\SI{16}{h}. Cells were harvested by centrifugation. For target protein purification and analysis, cell pellets were resuspended in lysis buffer and then sonicated. The supernatant after centrifugation was kept for future purification. BHRF1 was purified by one-step purification using strep column, while \gls{mdm2} and \gls{mcl1} were purified using Ni-affinity chromatography followed by size-exclusion chromatography (Superdex 75). Purified proteins were sterilized using a \SI{0.22}{\micro\meter} filter before being aliquoted for storage. The concentration of BHRF1 was determined by Bradford protein assay with BSA as standard. Protein purity and molecular weight were determined by standard SDS-PAGE and Western blot.

All other target proteins for \gls{mDisplay} were acquired commercially as described in \cref{tab:target-proteins}. 

\begin{table}[htbp]
\centering
\begin{tabular}{@{}lllll@{}}
\textbf{Target Protein} & \textbf{Vendor} & \textbf{cat. no.} & \textbf{Tags} & \textbf{Expression System} \\ 
\midrule
\gls{il7ra}(21-239)   & Bio-Techne      & 10758-IR               & C-terminal his-tag   & HEK293                    \\
\gls{pdl1}(19-239)    & Bio-Techne      & 9049-B7-100            & C-terminal his-tag   & HEK293                    \\
\gls{sc2rbd}(319-541)  & Sino Biological & 40592-V08B-B           & C-terminal his-tag   & Baculovirus-Insect Cells \\
\gls{trka}(34-423)     & Bio-Techne      & 9966-TK-050            & C-terminal his-tag   & HEK293                    \\
BHRF1(2-160)     & GenScript       & Custom Order\textsuperscript{+} & N-terminal his- and strep-tag& \textit{E. coli}          \\
\gls{mdm2}(1-188)     & GenScript       & Custom Order\textsuperscript{+} & N-terminal his-tag   & \textit{E. coli}          \\
\gls{mcl1}(172-327)     & GenScript       & Custom Order\textsuperscript{+} & N-terminal his-tag   & \textit{E. coli}          \\

\end{tabular}
\vspace{0.3cm}
\customtablecaption{\textbf{Commercially acquired target proteins for \gls{mDisplay}.} Ranges provided after the target protein names are amino acid residue ranges. \textsuperscript{+}Custom order fulfilled by GenScript, as detailed in \cref{app:Protein expression and purification}.}
\label{tab:target-proteins}
\end{table}

\subsection{Macrocycle cyclization}
\label{app:cyclization}

Peptides were synthesized by GenScript, as follows, using standard Fmoc-based solid-phase peptide synthesis on modified chloride resin. Amino acids were sequentially coupled to the resin with Fmoc deprotection steps using piperidine/DMF. The coupling and deprotection cycle was repeated until the full linear sequence was assembled, with reaction progress monitored by colorimetric resin tests.  Once the linear peptide was fully assembled, it was cleaved from the resin using a 1:3 (v/v) trifluoroethanol/dichloromethane solution, maintaining side-chain protection.

Cyclization was carried out in solution using PyBOP and DIPEA in DMF or DMF/DMSO (2:1 v/v) for 2--10 hours, followed by solvent removal via rotary evaporation. Global deprotection and final cleavage were achieved using TFA-based cocktails for 2--3 hours. The reaction mixture was precipitated into cold tert-butyl methyl ether, centrifuged, and the supernatant decanted to yield the crude macrocycle. The crude product was purified by preparative HPLC with a mobile phase containing 0.1\% TFA. Collected fractions were analysed by ESI-MS and analytical HPLC to confirm identity and purity. Fractions with \textgreater90\% purity were pooled and lyophilized to obtain final peptide powders.

\subsection{Mammalian display for detection of mini-binder binding}
\label{app:mdisplay}

A high-level overview of mammalian display is as follows (more details are provided in subsequent paragraphs). \textit{De novo} binder sequences are cloned into mDisplay vector. This vector enables targeting and anchoring of mini-binders to the cell surface while also expressing HA-tag. After transfection, displayed mini-binders are then exposed to target proteins with his-tag. Unbound target protein is washed away. Then, fluorescent antibodies bind to HA- or his-tags to indicating the presence of binder or target proteins (respectively) which is detected and quantified using high content imaging.

Binder amino acid sequences were codon optimized for expression in \textit{Homo sapiens} cells using Twist Bioscience's Codon Optimization tool \cite{twist_codon_tool} and synthesized and cloned into a modified version of pDisplay vector (pDisplay™ Mammalian Expression Vector, Invitrogen, cat. no. V66020) by Twist Bioscience \cite{ho2009mammalian}. This vector enables targeting and anchoring of proteins of interest to the cell surface, specifically, the \textit{de novo} binder is in an expression cassette containing an N-terminal secretion signal, HA-tag, myc-tag, and C-terminal transmembrane anchoring domain of platelet-derived growth factor receptor (PDGFR).

To prepare plasmids for transfection, glycerol stocks prepared by Twist were stamped onto LB+Ampicilin (\SI{100}{ug/mL}) agar plates (Teknova, cat. no. L1004) and incubated overnight at \SI{37}{\celsius}. This was then used to inoculate \SI{1.5}{mL} of LB+Carbenicillin (\SI{100}{ug/mL}) liquid and grown overnight at \SI{37}{\celsius} shaking at \SI{900}{RPM} for \SI{18}{hours}. Grown cultures were then pelleted at \SI{1000}{RPM} and used to purify plasmids by the University of California Berkeley DNA Sequencing Facility using the following procedures.

Pellets were resuspended in \SI{100}{\micro L} of Solution I (\SI{10}{mM} EDTA containing \SI{100}{ug/mL} RNAse A) by repeated robotic pipetting (Beckman Coulter BioMek FXp) and shaking on a plate mixer. Solution I was incubated for a total of \SI{5}{minutes} at room temperature (RT). To lyse the bacteria, \SI{100}{\micro L} of Solution II (\SI{0.2}{N} NaOH with \SI{1}{\percent} sodium dodecyl sulfate) was added and briefly mixed by repeated pipetting and shaking on a plate mixer, followed by incubation for a total of \SI{5}{minutes} at RT. Finally, \SI{100}{\micro L} of Solution III (\SI{3}{M} KOAc, pH5.2) was added and briefly mixed by pipetting, followed by shaking for \SI{10}{minutes} at RT to thoroughly mix. The culture block was subsequently centrifuged (\SI{4300}{g} for \SI{30}{min} at \SI{12}{\celsius}) to separate the lysate from the bacterial debris (which formed a pellet), and the block was placed back onto the FXp robot. \SI{10}{\micro L} of magnetic beads were aliquoted into a new 96-well round-bottom plate (Corning Costar 3799), which was placed onto the deck of the FXp. The FXp then removed \SI{140}{\micro L} of cleared lysate from the culture block, moved it to the new plate, and pipet mixed the lysate with the magnetic beads. Next, \SI{100}{\micro L} of isopropanol was added to precipitate the DNA onto the beads, and the plate was moved to a 96-well magnet for \SI{15}{minutes}. The DNA (on the beads) was rinsed 3 times for \SI{30}{seconds} with \SI{70}{\percent} ethanol. Ethanol was removed following the third rinse and the DNA was allowed to dry for \SI{15}{minutes}. Finally, \SI{30}{}\SI{50}{\micro L} of purified water was added in order to elute the DNA from the surface of the beads. To facilitate complete elution, the plate was sealed with plastic film and moved to a Thermomixer C (Eppendorf) with a heated cover set at \SI{60}{\celsius}, and was shaken at \SI{500}{RPM} for \SI{15}{minutes}. The plate was manually returned to a plate magnet, the eluates were removed to a clean 96-well polypropylene PCR plate, and \SI{2.5}{\micro L} of each eluted plasmid DNA was analysed for concentration and purity by absorbance at \SI{260}{nm}, \SI{280}{nm}, and \SI{320}{nm} light on a Bio-Tek Synergy plate reader.

To prepare for transfection, HEK293T cells (ATCC, cat. no. CRL-3216)  were cultured in Dulbecco's modified Eagle's medium with GLUTAMAX (Gibco, cat. no. 10569010), supplemented with \SI{10}{\percent} fetal bovine serum (Gibco, cat. no. A5670502), and 1X Penicillin/Streptomycin (Gibco, cat. no. 15140122). Cells were grown at \SI{37}{\celsius} with \SI{5}{\percent} CO2 in a standard tissue culture incubator and passaged at 70-\SI{80}{\percent} confluency using TrypLE Express (Gibco, cat. no. 12604039). 

To transfect, \SI{210}{ng} of purified clonal plasmid DNA containing the binder was complexed with TransIT®-293 (Mirius Bio, cat. no. MIR 2704) and placed into a single well of a Poly-D-lysine (Gibco, cat. no. A3890401)-coated Revvity 96 Well Plate (Revvity, cat. no. 6005225). \num{20000} HEK293T cells, passage 11, were then added. After \SI{48}{hours} of incubation at \SI{37}{\celsius} with \SI{5}{\percent} CO2, cells were incubated with \SI{30}{\micro L} primary binding solution (OptiMEM, Gibco, cat. no. 31985062, \SI{3}{\percent} fetal bovine serum, \SI{0.02}{mg/mL} target protein, Table S1) for \SI{30}{minutes} at \SI{37}{\celsius} with \SI{5}{\percent} CO2.  After primary binding, cells were washed once with warm \SI{100}{\micro L} DPBS containing calcium and magnesium (Gibco, cat. no. 14040133). Cells were then incubated with \SI{30}{\micro L} secondary binding solution (OptiMEM, \SI{3}{\percent} fetal bovine serum, 1:200 CoraLite® 594-conjugated 6*His, his-tag Mouse McAb, Proteintech, cat. no. CL594-66005, 1:200 CoraLite® Plus 488-conjugated HA tag Mouse McAb, Proteintech, cat. no. CL488-66006) for \SI{40}{minutes} at \SI{37}{\celsius}. Cells were washed once with warm \SI{100}{\micro L} DPBS containing calcium and magnesium. For imaging, cells were kept in imaging solution (OptiMEM, no phenol-red, Gibco, cat. no. 11058021, \SI{3}{\percent} fetal bovine serum). 

High content imaging was performed on an Operetta CLS High-Content Analysis System (Perkin Elmer) using a 10X Air NA 0.3 confocal. Example images can be found in \subfigref{fig:Definition of mDisplay}{f}, of a positive control binder in two wells, either with or without exposure to target protein as labelled. Prior to use, Operetta was set to \SI{37}{\celsius}. Cells were identified and analysed using digital phase contrast (Harmony, Method P), and mean fluorescence intensity for both channels were quantified to detect binder expression (488) and target binding (594), respectively, per identified cell. Wells containing fewer than \num{2000} identified cells were excluded from analysis, except for BHRF1, where a relaxed threshold of 500 cells was applied. Non-transfected cells were excluded by gating for sufficient 488 (binder) fluorescence above the 95th percentile of an untransfected negative control well (confluent, non-expressing; \subfigref{fig:Definition of mDisplay}{a}). \subfigref{fig:Definition of mDisplay}{c}, \subfigref{fig:Definition of mDisplay}{d}, and \subfigref{fig:Definition of mDisplay}{e} are selected per-well results for one positive control (RFD\_PDL1\_76) and two \textit{de novo} binders (LL\_MINI\_PD-L1\_39 and LL\_MINI\_PD-L1\_36) with differing observed binding strengths. For each well, mean \SI{594}{nm} fluorescence intensity (MFI) was calculated and normalized to the MFI of a plate-matched, positive control binder without target exposure  (\subfigref{fig:Definition of mDisplay}{b}; \cref{tab:positive-binders}) to correct for plate effects and possible affinity variation across targets' positive control binders, described in \cref{tab:positive-binders}. This value is named ‘‘Relative MFI’’ as seen in \cref{fig:Definition of mDisplay}, \cref{fig:mammalian HT-BLI correlation}, and \cref{fig:specificity_assay}.

For each \denovo{} design, the mean normalized MFI across replicate wells was computed. Designs with MFI $\leq 1$ were excluded from further analysis. Remaining designs were evaluated via two-sided independent t-tests, comparing their MFI distributions to those of pooled negative controls within the same target. Resulting $p$-values were corrected for multiple hypothesis testing using the Benjamini–Hochberg procedure. Designs with FDR-adjusted $p$-values (or $q$-values) <0.05 were considered significant for \subfigref{fig:mammalian HT-BLI correlation}{a}.

\subsection{BLI}
\label{app:BLI}
All \gls{BLI} assays were performed by GenScript using a Sartorius Octet RED384 system with corresponding biosensors at \SI{30}{\celsius} and \SI{1000}{RPM}. Assays were conducted in black polypropylene flat-bottom assay plates (Greiner, cat. no. 5085651) using PBST buffer (\SI{0.03}{\percent} Tween-20 in PBS, pH 7.2; GenScript).
For the \gls{HT-BLI} assay, biosensors were hydrated in PBST for \SI{20}{minutes} at room temperature and conditioned by three cycles of \SI{5}{second} immersion in regeneration buffer followed by \SI{5}{second} immersion in PBST (neutralization buffer). Sensors were equilibrated in PBST for \SI{60}{seconds} to establish a baseline.

Mini-binders were diluted 10-fold in PBST to achieve uniform loading. Sensors were immersed in the cell-free expression supernatant to load the protein to saturation (\SI{0.5}{nm} shift), followed by equilibration in PBST for \SI{200}{seconds}. Association was measured by immersing sensors in PBST containing \SI{1000}{nM} antigen for \SI{150}{seconds}. Dissociation was monitored by transferring sensors back to PBST for \SI{30}{seconds}. A zero-concentration analyte (PBST buffer alone) was used as a reference.

Mini-binder designs with HT-BLI response >\SI{0.03}{RU} after reference subtraction were classified as designs with measurable binding signal. Specifically, responses <\SI{0.03}{RU} were considered non-binders, 0.03--\SI{0.10}{RU} intermediate, and >\SI{0.10}{RU} strong.
For the 5-concentration \gls{BLI} Assay, biosensors were hydrated and conditioned as described above. Proteins were diluted to appropriate concentrations in PBST for immobilization. Sensors were immersed in the antibody solution and loaded to ~\SI{1}{nm} shift, followed by equilibration in PBST for \SI{300}{seconds}. Antigen was serially two-fold diluted in PBST to generate five concentrations. Association was measured by immersing sensors in each antigen concentration solution, and dissociation was monitored by transferring sensors back to PBST. A zero-concentration analyte (PBST buffer alone) was included as a reference.

Control and reference binders RFD\_PDL1\_76, RFD\_TrkA\_88, and GDM\_TrkA\_9 exhibited low immobilization levels even at saturating immobilization concentrations and did not produce response in 5-concentration BLI. Upon further characterization using 5-concentration SPR, \kd{} was derived for RFD\_PDL1\_76 but not RFD\_TrkA\_88, as described in \cref{app:benchmark}.

All data were processed using Octet RED BLI Discovery software version 12.2.2.26. In preliminary experiments, non-specific binding was tested to determine appropriate experimental conditions for subsequent analyses.

\subsection{SPR}
\label{app:SPR}
All \gls{SPR} assays were performed by GenScript using a Biacore 8K system (Cytiva) at \SI{25}{\celsius}. Series S CM5 sensor chips were used for all assays. Target proteins were immobilized by amide coupling using the Amine coupling Kit (Cytiva). 
Binding measurements were conducted using either multi-cycle kinetics or single-cycle kinetics, depending on the specific interaction characteristics of the analyte and ligand.

Analytes were injected as 8-point or 5-point concentration series, with starting concentrations ranging from \SI{10}{\micro M} to \SI{1000}{\micro M} depending on the sample. Injections were performed at a flow rate of \SI{30}{\micro L/min}, with an association time of \SI{60}{seconds} and a dissociation time of \SI{60}{seconds}. Running buffers were selected based on peptide/protein solubility and formulation requirements, these included ultrapure water, DMSO, and PBS buffer.
All data was processed using Biacore 8K Evaluation Software (version 5.0). Sensorgrams were double-referenced using a blank reference surface and buffer-only injections to correct for nonspecific binding and bulk refractive index effects. We used two fitting approaches to determine binding affinity: kinetic fitting and steady-state fitting. The choice of method depended on the interaction characteristics between the analyte and ligand. For interactions with clearly measurable association and dissociation phases, kinetic analysis was performed using multi-cycle kinetics. For fast-on/fast-off interactions, steady-state fitting was used instead, based on equilibrium binding responses.

\end{document}